\RequirePackage{fix-cm}
\documentclass[smallextended]{svjour3}       
\smartqed  
\usepackage{graphicx}
\usepackage{natbib}
\usepackage[monochrome]{color}
\journalname{Space Science Reviews}
\begin{document}

\title{Noble gases and stable isotopes track the origin and early evolution of the Venus atmosphere
}

\titlerunning{Venus: noble gases and stable isotopes}        

\author{Guillaume Avice         \and
        Rita Parai \and Seth Jacobson \and Jabrane Labidi \and Melissa G. Trainer \and Mihail P. Petkov
}

\institute{G. Avice \at
              Université Paris Cité, Institut de physique du globe de Paris, CNRS, 75005 Paris, France \\
              Tel.: +33-183957731\\
              \email{avice@ipgp.fr}           
           \and
           R. Parai \at
              Department of Earth and Planetary Sciences and McDonnell Center for Space Sciences, Washington University in St. Louis, St. Louis, MO, USA\\
              Tel.: +1-314-935-3974\\
              \email{parai@wustl.edu}
              \and
           S. A. Jacobson \at
              Department of Earth and Environmental Sciences, Michigan State University, East Lansing, MI, USA 4886\\
              \email{seth@msu.edu}
              \and
           J. Labidi \at
              Université Paris Cité, Institut de physique du globe de Paris, CNRS, 75005 Paris, France \\
              \email{labidi@ipgp.fr}
              \and
           M. Petkov \at
              NASA Jet Propulsion Laboratory, California Institute of Technology, Pasadena, CA 91109 USA\\
              \email{mihail.p.petkov@jpl.nasa.gov}
              \and
           M. G. Trainer \at
              NASA Goddard Space Flight Center, Greenbelt, MD 20771, USA\\
              \email{melissa.trainer@nasa.gov}
}

\date{Received: date / Accepted: date}

\maketitle

\begin{abstract}
The composition the atmosphere of Venus results from the integration of many processes entering into play over the entire geological history of the planet. Determining the elemental abundances and isotopic ratios of noble gases (He, Ne, Ar, Kr, Xe) and stable isotopes (H, C, N, O, S) in the Venus atmosphere is a high priority scientific target since it could open a window on the origin and early evolution of the entire planet. 
This chapter provides an \textcolor{red}{overview} of the existing dataset on noble gases and stable isotopes in the Venus atmosphere.
The current state of knowledge on the origin and early and long-term evolution of the Venus atmosphere deduced from this dataset is summarized. 
A list of persistent and new unsolved scientific questions \textcolor{red}{stemming from recent studies of planetary atmospheres (Venus, Earth and Mars)} are described. Important mission requirements pertaining to the measurement of volatile elements in the atmosphere of Venus as well as potential technical difficulties are outlined.


\end{abstract}

\section{Introduction and Overview}\label{sec:introduction}
Like all planetary atmospheres, the composition of the atmosphere of Venus informs us about the entire geological history of Earth's sister planet \citep[\textit{e.g.}][]{catling_atmospheric_2017}. Past space missions have already demonstrated that the elemental and isotopic compositions of volatile elements in the atmosphere of Venus differ greatly from those of other terrestrial planets or reservoirs of volatile elements (solar gas, meteorites, comets) in the Solar System \citep[\textit{e.g.}][]{pepin_origin_2002,baines_atmospheres_2013,wieler_noble_2002,chassefiere_evolution_2012,avice_perspectives_2020}. Recent discoveries about the detailed origins and evolution of volatile inventories on Earth, Mars, and in other Solar System reservoirs have raised compelling new questions that motivate the pursuit of improved noble gas and stable isotope measurements in the atmosphere of Venus. Future space missions targeting atmospheric measurements are needed to make critical comparisons of volatile accretion and transport between terrestrial planets.

Noble gases are the best available geochemical tracers of geophysical processes. Noble gases are inert, incompatible during silicate partial melting, and atmophile \citep{ozima_noble_2002}. Accordingly, noble gases act as tracers of planetary differentiation, including the coupled evolution of planetary interiors and atmospheres through outgassing. Each noble gas element has at least one stable isotope that is non-radiogenic; these are often referred to as primordial noble gas isotopes, and their budgets are established during accretion. Ratios of primordial noble gas isotopes serve as fingerprints of volatile origins, but these ratios may also be affected by mass-fractionating loss. Each noble gas element also has at least one stable isotope that is produced by nuclear reactions (including radioactive decay) involving non-atmophile reactants. Noble gas abundances are typically so low in planetary solids that their isotopic compositions are very strongly affected by nuclear reactions, even when the reactants are themselves rare in planetary materials. Thus, radioactive decay and other nuclear reactions generate large variations (from a few percent to orders of magnitude) that track planetary differentiation, which fractionates non-atmophile parent to atmophile daughter ratios, on a variety of timescales. Taken together, noble gas elemental abundances and isotopic compositions provide a rich record of planetary volatile origins, and the timing and mechanisms of volatile transport between planetary reservoirs.

Stable isotope ratios of relatively light elements (C, H, O, N, S) have been widely used to put constraints on the budget of volatile elements of terrestrial planets and on the origin and evolution of planetary atmospheres \cite[\textit{e.g.}][]{marty_origins_2012,halliday_origins_2013}. These elements do not show the strong nuclear effects seen for noble gases. Nonetheless, various reservoirs in the Solar System show variable mass-dependant and mass-independent isotopic composition. In solar-system materials, isotope variations may have been caused by physico-chemical processes such as photochemistry, partial melting, and partial evaporation. Planetary formation and differentiation may lead to further isotopic variations \citep{young_kinetic_2002}. The specific case of atmospheric loss may also lead to isotopic fractionation \citep{jakosky_mars_1994}. In other words, the isotope ratios of light elements can help track the origin of planetary materials but also place constraints on the evolution of planetary mantles and atmospheres.

The possible sources of Venusian atmophile elements (noble gases and light elements) can be categorized roughly into two types: direct or indirect.
Venus could have accreted a significant primary atmosphere \textbf{directly} from the nebula if it accreted as quickly as Mars \citep{dauphas_hfwth_2011} and reached near its final size while gas was likely still present in the protoplanetary disk \citep{weiss_history_2021}.
In this case, the isotopic composition of atmophile elements sourced directly from the solar nebula would presumably match those of gases of the proto-Sun.
While hydrodynamic escape assisted by later bombardment from small bodies could have eroded away hydrogen and helium, other gases directly accreted from the nebula may have been in-gassed into the mantle of Venus and protected or may have survived in the atmosphere \citep{olson_nebular_2019}. 
However, if Venus accreted more slowly and was Mars-sized or smaller only growing to larger sizes after the gaseous protoplanetary disk had dissipated, one may anticipate that a primary atmosphere would have been insignificant.
Regardless of the size of its primary atmosphere, Venus also probably accreted volatile-rich building blocks during its growth providing an \textbf{indirect} source of atmophile elements.
The isotopic composition of atmophile elements sourced indirectly from accreted building blocks would vary in composition according to the parent-body source region as well as fractionating processes taking place within the parent-body.
During the dynamical evolution of the solar system, different reservoirs of planetary building blocks contributed to the growth of the terrestrial planets \citep{obrien_delivery_2018}.
Initial accretion in the inner disk near the current location of Venus is presumed to be from volatile-depleted material, although even enstatite and ordinary chondrites, examples of primordial inner solar system material, contain substantial amounts of volatile species like water \citep[e.g.][]{piani_earths_2020}.
If not supplied in Venus's building blocks, most volatiles within Venus could have been delivered from more distant parts of the protoplanetary disk, where volatile abundances in solids were significantly higher.
However, this region is vast, encompassing primitive asteroidal and cometary compositions with clear evidence for strong isotopic gradients throughout \citep{marty_origins_2016}.
Only with a detailed understanding of the isotopic composition of the venusian atmosphere will we be able to disentangle the different contributions to the atmosphere of Venus.

Apart from remote observations, and compared to missions involving landing on the surface of Venus, an exhaustive measurement of the elemental and isotopic composition of volatile elements in a sample of Venus's atmosphere is one of the less risky types of investigations with an astounding scientific outcome. Measuring the elemental and isotopic composition of noble gases and stable isotopes in the atmosphere of Venus is for example one of the main scientific goals of the recently selected NASA DAVINCI mission \citep{glaze_davinci_2017}.

This chapter presents an overview of the current state of knowledge and the outlook for future discoveries brought from the study of noble gases and of stable isotopes of C, H, O, N and S regarding the origin and early evolution of the atmosphere of Venus and on the implications for the geological history of the entire planet. Note that other contributions in this issue will cover in detail the role of water in Venus's history \citep{salvador_magma_2022} and the complex interactions between the interior and the surface of Venus \citep{gillmann_long-term_2022}.

\section{\bf Existing dataset and current state of knowledge}
The current state of knowledge on the elemental and isotopic compositions of noble gases and stable isotopes of H, C, N and O in the Venus atmosphere \textcolor{red}{has already been reviewed in details elsewhere \citep{von_zahn_composition_1983,donahue_origin_1983,wieler_noble_2002,johnson_venus_2019,chassefiere_evolution_2012}}. This section describes the most important features of the atmosphere of Venus and highlights which scientific data are severely lacking.

\subsection{Noble gases (He, Ne, Ar, Kr, Xe)}\label{subsec:datanoblegases}
Existing data on the elemental abundance and isotopic composition of noble gases in the atmosphere of Venus are reported in Table \ref{tab:NobleGases}.\,

\begin{table}
\caption{\textcolor{red}{TABLE MODIFIED DURING THE REVISION.} Noble gases in the atmosphere of Venus. Abundances are reported as molar mixing ratios in the atmosphere. See details in the text for estimates for the abundance of xenon. Values in blue font give the percentages for errors. \textcolor{red}{Recommended} precisions for future investigations are taken from \citet{chassefiere_evolution_2012}. Table modified after \citet{wieler_noble_2002}.}
\label{tab:NobleGases}       
\includegraphics[width=1\textwidth]{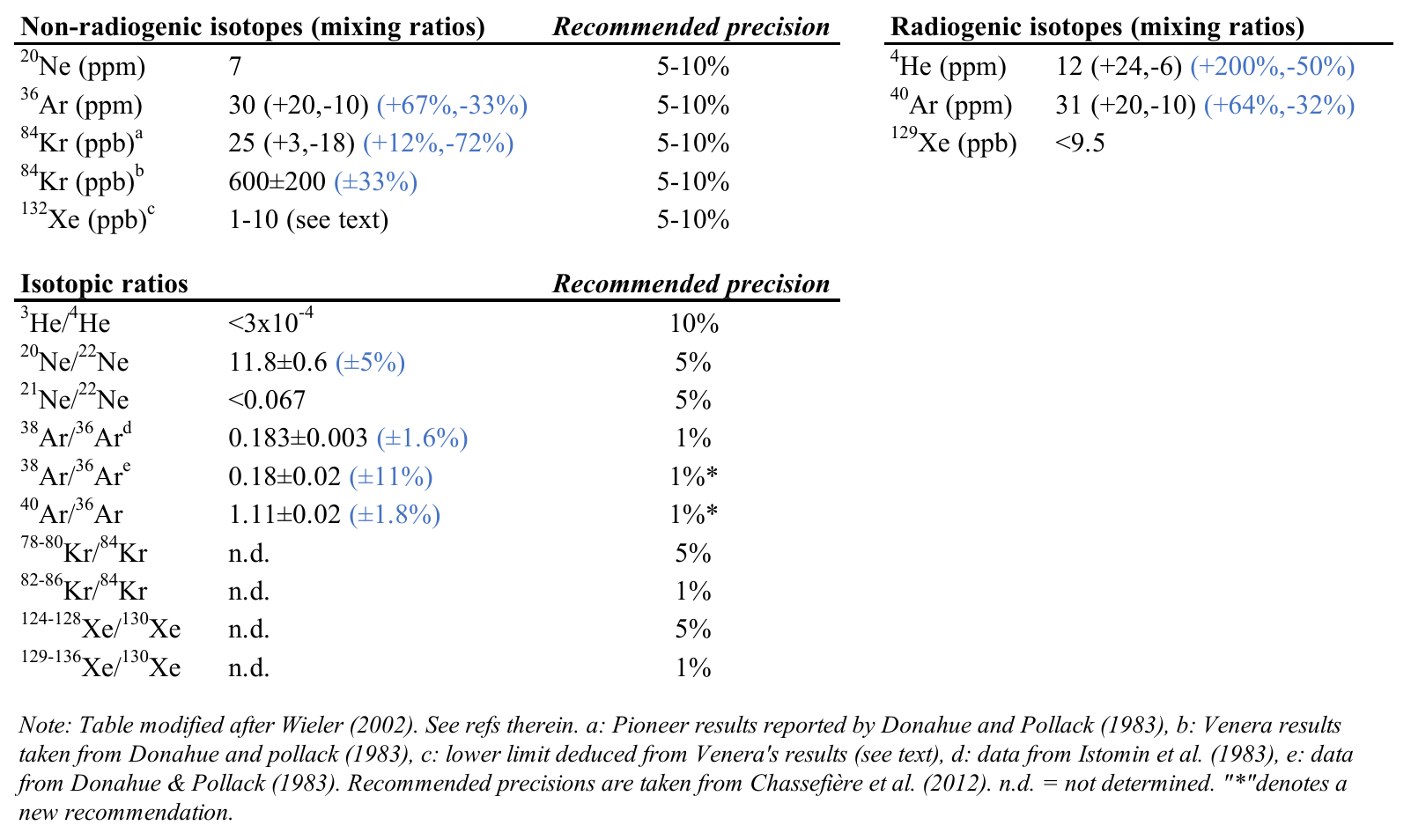}
\end{table}

\subsubsection{Elemental abundances}
\begin{figure*}
\centering
  \includegraphics[width=0.75\textwidth]{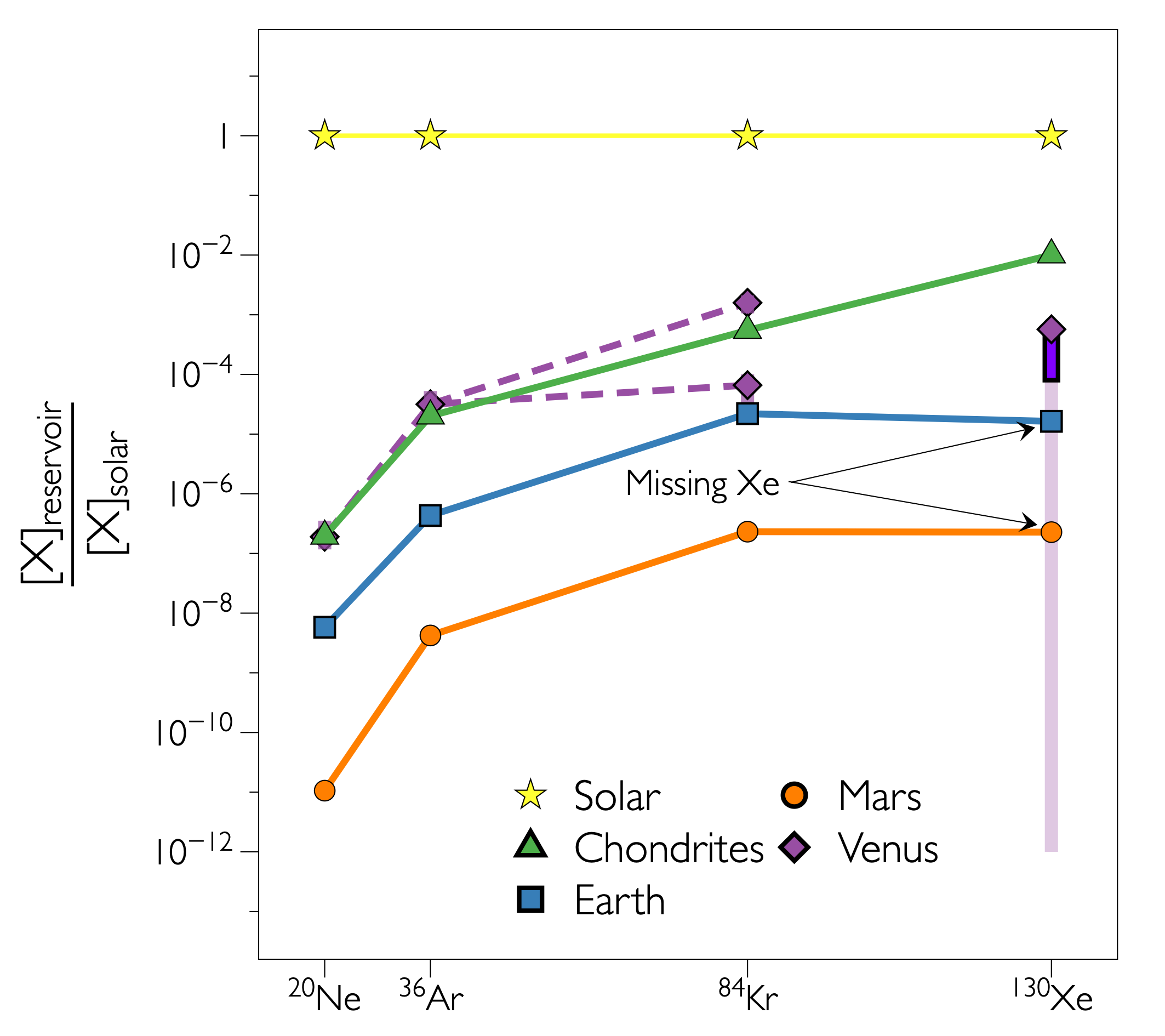}
\caption{\textcolor{red}{FIGURE MODIFIED DURING THE REVISION.} Elemental abundances of noble gases in the atmosphere of Venus and in other reservoirs of the Solar System normalized to the abundance of silicium and to the Solar abundances (=1). For Venus, uncertainties on the abundances of Kr and Xe appear with a light purple range (this is an upper limit for Xe). The dark purple range for $^{130}$Xe corresponds to another estimate with a minimum of 1 ppb of Xe (see text). The dotted purple line depicts the expected abundances of $^{130}$Xe for an Earth-like/Mars-like $^{130}$Xe/$^{84}$Kr ratio. 
Figure and data modified after \citet{dauphas_geochemical_2014}. See also references therein.}
\label{fig:AbundancesFromDauphas}       
\end{figure*}

\begin{figure*}
\centering
  \includegraphics[width=0.75\textwidth]{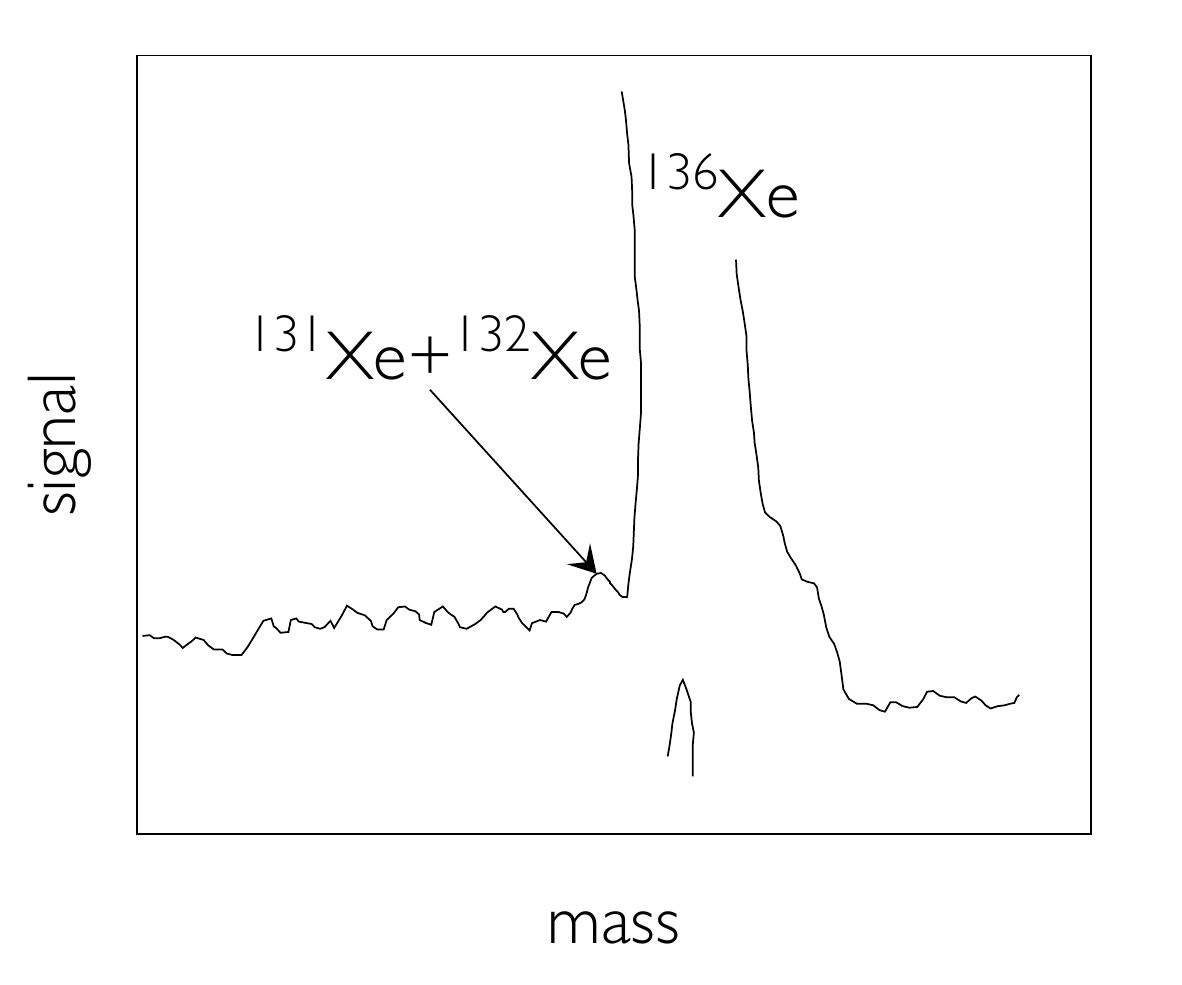}
\caption{Signal of $^{131+132}$Xe detected during the Venera 14 mission. Mass and signal are given without units and the signal at the top of the $^{136}$Xe peak is divided by 100. The very high $^{136}$Xe peak is due to the presence of calibrant gas (99.99\% pure $^{136}$Xe). The $^{131+132}$Xe could correspond to a minimum abundance of 1-2 ppb of Xe in the atmosphere of Venus. Figure modified after \citet{istomin_preliminary_1982}.}
\label{fig:IstominDetection}       
\end{figure*}

Elemental abundances of Ne, Ar, Kr and Xe in the atmosphere of Venus and in other reservoirs of volatile elements in the Solar System are plotted in Fig.\,\ref{fig:AbundancesFromDauphas} \citep{dauphas_geochemical_2014,avice_perspectives_2020}. \textcolor{red}{Chondritic meteorites, especially carbonaceous chondrites,} are enriched in heavy noble gases relative to light ones \citep{porcelli_models_2002,wieler_noble_2002}. Although Mars is depleted in noble gases relative to Earth by about two orders of magnitude, note that noble gases in these two atmospheres follow globally the same abundance pattern with a similar magnitude of enrichment of Kr relative to Ne relative to Ar. Atmospheric xenon on Earth and Mars presents a depletion relative to a chondritic-like abundance pattern (this is the "missing xenon problem," see \citet{avice_perspectives_2020} for a recent review).
\textcolor{red}{For Venus, abundances of Ne and Ar are high compared to Mars and Earth and are similar to those measured in meteorites. Two ranges of estimates exist for the abundance of krypton (see Table \ref{tab:NobleGases}, \citet{donahue_origin_1983} and refs. therein). If there is 25 ppb of $^{84}$Kr in the Venus's atmsphere (Pioneer's results), 
krypton would be depleted relative to a chondritic-like abundance pattern, with almost a solar-like $^{84}$Kr/$^{36}$Ar ratio. Conversely, if the abundance is closer to 600 ppb (Venera's results), the $^{84}$Kr/$^{36}$Ar ratio is close to the chondritic ratio (Fig. \ref{fig:AbundancesFromDauphas}), suggesting that Ne, Ar and Kr in the Venus's atmosphere could have been sourced by meteorites. Measuring precisely the abundance of $^{84}$Kr would be decisive for understanding the origin of volatile elements in Venus.}
For xenon, an upper limit of 10\,ppb for the abundance of $^{132}$Xe \textcolor{red}{has often been cited} based on results obtained during the Pioneer mission \citep[e.g.][]{wieler_noble_2002}. However, \citet{istomin_preliminary_1982,istomin_venera_1983} argued that xenon was detected during the Russian Venera 14 mission and that, according to the sensitivity of the instrument, a detection of $^{131+132}$Xe isotopes above background implies a minimum abundance of Xe of 1-2\,ppb (Fig. \ref{fig:IstominDetection}). A new precise determination of the abundance of Xe is still pending but considering a 1-10 ppb range suggests that Xe is not more depleted in the Venus atmosphere than atmospheric Xe on Earth and Mars. The fact that Xe in the atmospheres of Earth, Mars and Venus could be depleted in a similar fashion compared to chondritic abundances is striking, especially if the underabundance is due to a selective escape mechanism of xenon ions coupled to hydrogen ions during hydrodynamic escape of hydrogen \citep{zahnle_strange_2019,avice_perspectives_2020}. 

\subsubsection{Isotope ratios}

\begin{figure*}
\centering
  \includegraphics[width=0.75\textwidth]{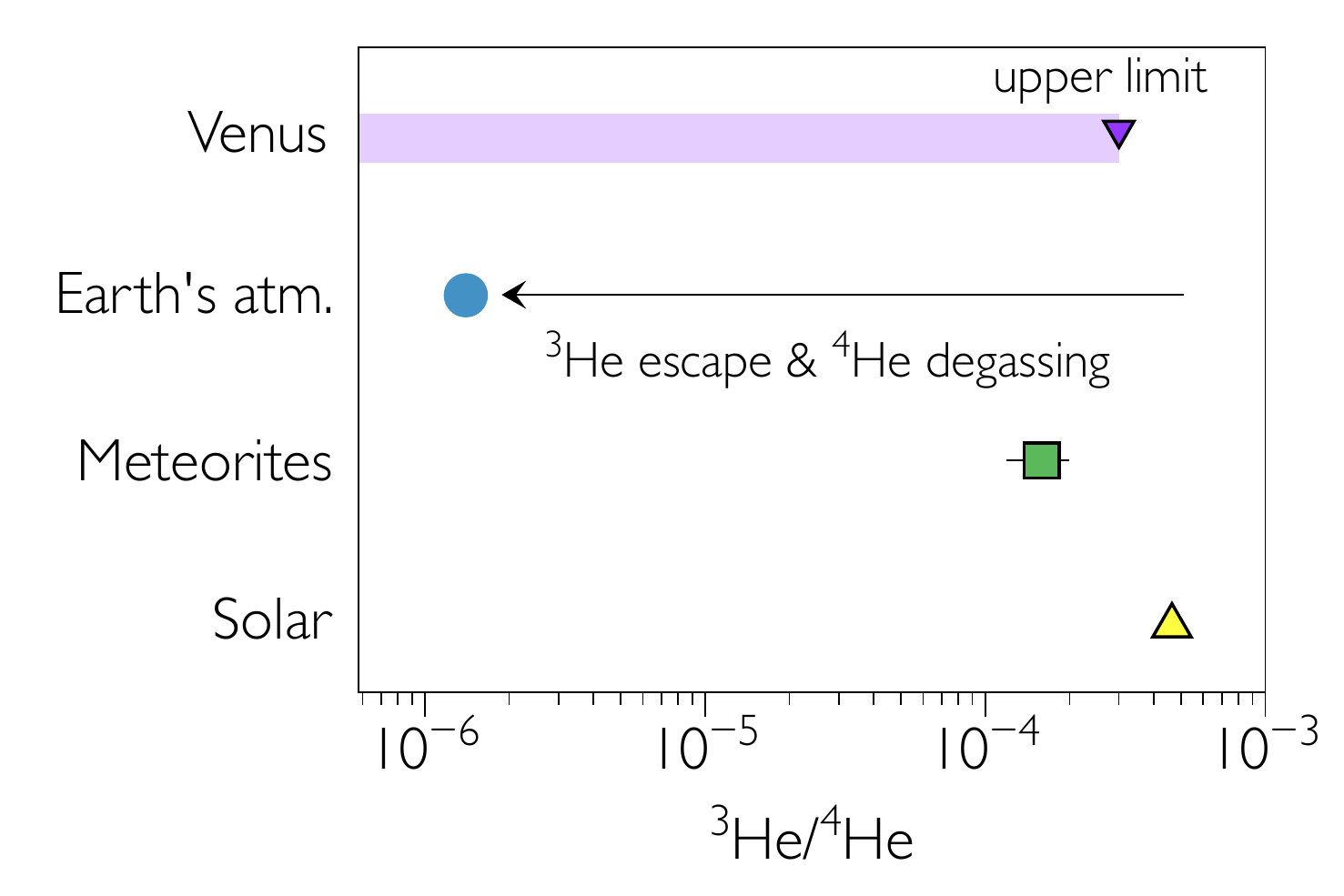}
\caption{$^{3}$He/$^{4}$He ratio of the atmosphere of Venus and of other reservoirs in the Solar System. Atmospheric escape of helium or degassing of radiogenic $^{4}$He leads to a decrease of the $^{3}$He/$^{4}$He ratio. Values are from \citet{porcelli_overview_2002} and refs. therein.}
\label{fig:Helium34}       
\end{figure*}

\textit{Helium}\,\textendash\, An upper limit for the $^{3}$He/$^{4}$He ratio of atmospheric helium has been estimated at $3\times10^{-4}$ \citep{hoffman_composition_1980}. This leaves a wide range of potential helium isotopic compositions viable for the atmosphere of Venus. The rather imprecise determination could be due to the scarcity of $^{3}$He relative to $^{4}$He, to the presence of an isobaric interference due to the presence of $^{1}$H$^{2}$D, or could be explained by the fact that droplets of sulfuric acid clogged the inlet port of the Pioneer probe, which interfered with measurements of noble gases \citep{hoffman_composition_1980}. This upper limit is compared to the $^{3}$He/$^{4}$He ratio of other reservoirs in the Solar System in Fig. \ref{fig:Helium34}. The atmospheric $^{3}$He/$^{4}$He ratio for Venus \textcolor{red}{seems only marginally lower} than the solar value, but since even the order of magnitude of the exact value remains unknown, it is impossible to evaluate the roles of relatively recent (hundreds of millions of years) outgassing of radiogenic $^{4}$He and of preferential escape of $^{3}$He relative to $^{4}$He from the atmosphere of Venus to space.   
\smallskip

\begin{figure*}
\centering
  \includegraphics[width=0.75\textwidth]{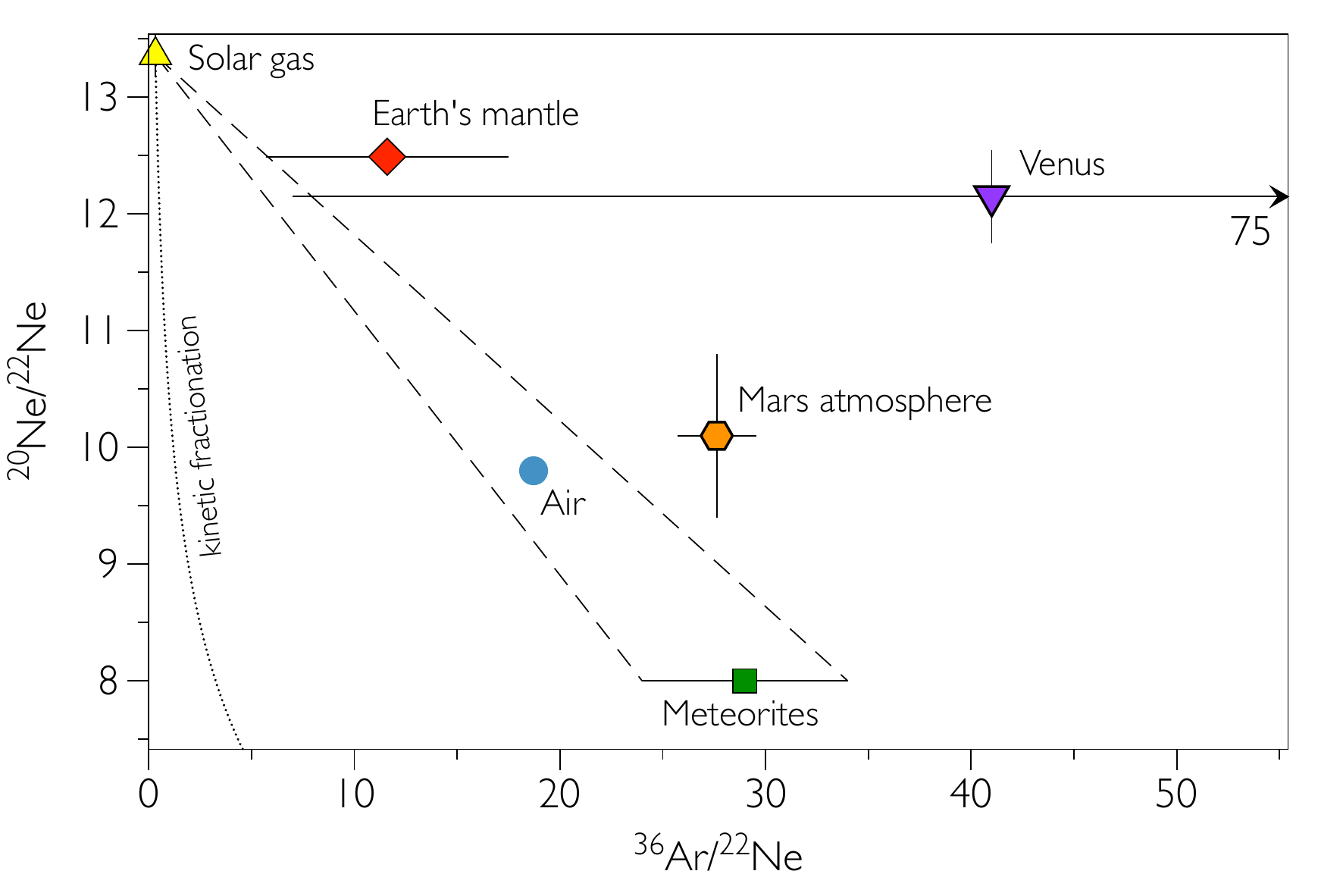}
\caption{Three-isotope diagram of neon isotopes ($^{20,21,22}$Ne). The isotopic composition of Earth atmospheric neon lies on a mixing range between Solar and Meteoritic end-members. The current estimate for Venus atmospheric neon suggests that, similarly to Mars atmospheric neon, Venus could lie outside of this mixing range although the uncertainty on the Ar/Ne ratio in the Venus atmosphere is high.  Figure modified after \citet{marty_origins_2012}.}
\label{fig:NeonIsotopes}       
\end{figure*}

\noindent
\textit{Neon}\,\textendash\,
The $^{20}$Ne/$^{22}$Ne ratio in the Venus atmosphere is close to 12 (Table \ref{tab:NobleGases}, Fig. \ref{fig:NeonIsotopes}). This ratio is higher than that measured for terrestrial atmospheric neon (9.8, \citet{ozima_noble_2002}), which suggests that solar-wind irradiated material ($^{20}Ne/^{22}Ne=12.8$, \citet{heber_noble_2009}) or solar gas ($^{20}Ne/^{22}Ne=13.4$, \citet{heber_isotopic_2012}) contributed volatile elements in greater proportion to the atmosphere of Venus relative to Earth. For the $^{21}$Ne/$^{22}$Ne ratio, only an upper limit of 0.067 has been proposed by \citet{istomin_venera_1983}. This value is too imprecise to provide any further constraint on the origin of neon on Venus.

\begin{figure*}
\centering
  \includegraphics[width=0.75\textwidth]{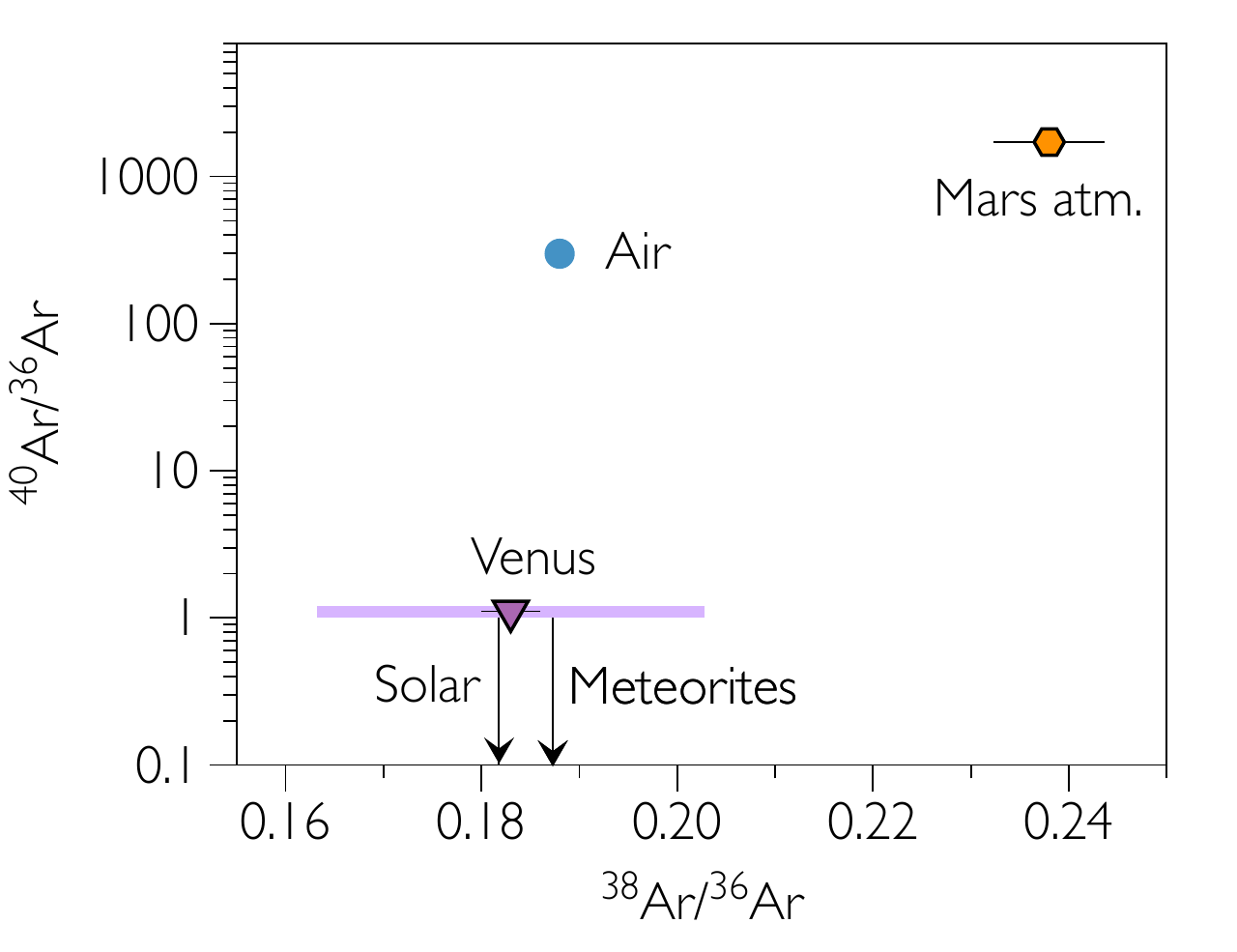}
\caption{Three-isotope diagram of argon isotopes. The thin black line corresponds to the error on the $^{38}$Ar/$^{36}$Ar ratio of Venus reported by \citet{istomin_venera_1983}. The purple range represents the uncertainty given by \citet{donahue_origin_1983}. Values for air are from \citet{ozima_noble_2002}. Values for Mars are taken from \citet{avice_noble_2018} and refs. therein. Error bars at 1$\sigma$.}
\label{fig:ArgonIsotopes}       
\end{figure*}

\smallskip
\noindent
\textit{Argon}\,\textendash\, There are two estimates for the $^{38}$Ar/$^{36}$Ar ratio, one with a value of 0.183$\pm$0.003 \citep{istomin_venera_1983}, and another with a \textcolor{red}{much less precise value of 0.18$\pm$0.02 \citep{donahue_origin_1983}. These two values are plotted in Fig. \ref{fig:ArgonIsotopes}. If the conservative error estimate is considered, the $^{38}$Ar/$^{36}$Ar ratio could correspond to any primordial end-member of the Solar System. However, if the value taken for the $^{38}$Ar/$^{36}$Ar ratio is 0.183±0.003, then the primordial isotopes of argon are similar to Solar argon.} The $^{40}$Ar/$^{36}$Ar ratio has been measured at 1.11$\pm$0.02 \citep{istomin_venera_1983}. Such a low value compared to the isotopic ratio of Earth's atmospheric argon ($\approx$300, \citet{ozima_noble_2002}) has been interpreted as evidence that Venus outgassed only about 25\% of the radiogenic $^{40}$Ar produced by the radioactive decay of $^{40}$K (T$_{1/2}$=1.25 Ga) during 4.56 Ga \citep{kaula_constraints_1999}. This \textcolor{red}{estimate for the amount of degassed radiogenic argon is about half the value estimated for Earth \citep{allegre_rare_1987}. Note that estimates for the total outgassing of $^{40}$Ar over Venus's history strongly rely on the assumed K/U ratio for bulk Venus, which remains debated (see the discussions by \citet{lammer_constraining_2020} and by \citet{gillmann_long-term_2022}).}
Limited outgassing of radiogenic Ar over 4.56 Ga is the basis for the common view that Venus remained quiescent, in a stagnant-lid tectonic regime, during most of its history \citep{orourke_thermal_2015}. \textcolor{red}{The high abundance of $^{36}$Ar means that a large fraction of argon in the Venus atmosphere (Fig.\,\ref{fig:AbundancesFromDauphas})} could be derived from primordial cosmochemical sources (solar gas, meteorites, comets). However, the relative abundances and isotope ratios of Ne and Ar do not provide a simple view of the source of light noble gases to Venus. The primordial Ar/Ne ratio is similar to values measured in meteorites, but the $^{20}$Ne/$^{22}$Ne ratio \textcolor{red}{may be} solar-like (see above). The large uncertainty on the determination of the elemental Ar/Ne ratio does not permit evaluation of whether Venus plots on the mixing line between solar and meteoritic end-members (Fig. \ref{fig:NeonIsotopes}), or if the delivery of gas with an extremely high Ar/Ne ratio (\textit{e.g.} via comets, \citet{marty_origins_2016}) induced an excess of primordial Ar relative to Ne without changing the $^{20}$Ne/$^{22}$Ne ratio.

\begin{figure*}
\centering
  \includegraphics[width=0.75\textwidth]{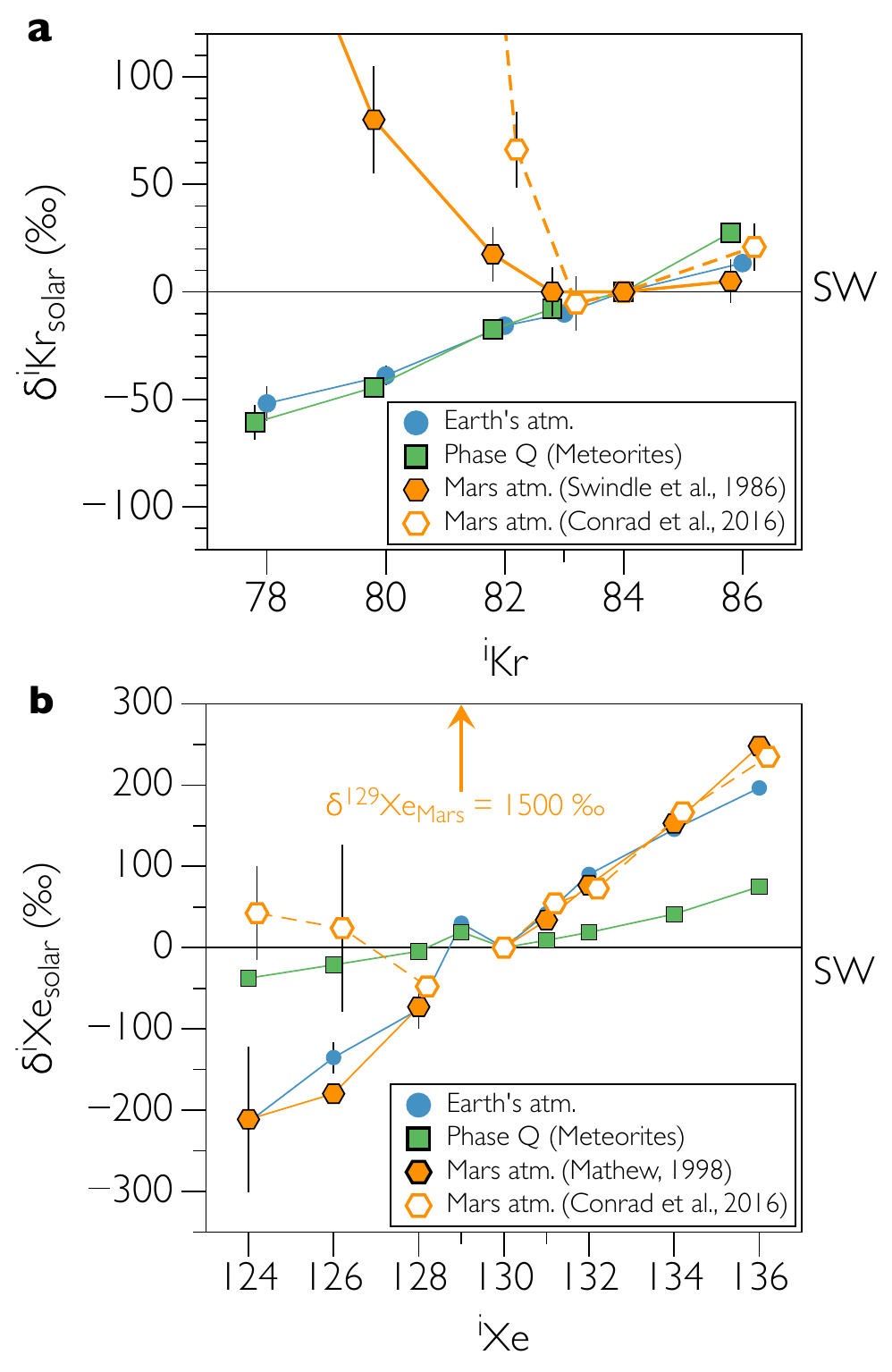}
\caption{\textcolor{red}{FIGURE MODIFIED DURING THE REVISION.} Isotopic composition of krypton (a) and xenon (b) in important reservoirs of the Solar system. Isotopic ratios are expressed in delta values, which correspond to deviations in permil relative to the isotopic composition of the Solar Wind \citep{meshik_heavy_2014,meshik_refined_2020}. Data for the Earth's atmosphere are from \citet{ozima_noble_2002}.
Data for Mars atmospheric Kr are from \citet{conrad_situ_2016} for in-situ measurements and from \citet{swindle_xenon_1986} for measurements made on Martian meteorites. Data for Mars atmospheric Xe are from \citet{conrad_situ_2016} for in-situ measurements and from \citet{mathew_martian_1998} for Xe measured in Martian meteorites.  
Note that the exact values, especially for light Kr and Xe isotopes, remain debated at this time \citep{avice_noble_2018}.
Data for Kr and Xe in the phase Q, the major carrier of heavy noble gases in carbonaceous chondrites, are from \citet{busemann_primordial_2000-1}. Error bars at 1$\sigma$.}
\label{fig:KryptonXenonIsotopes}       
\end{figure*}

\smallskip
\noindent
\textit{Krypton and Xenon}\,\textendash\, Today, there is no available measurement of the isotopic composition of krypton or xenon in the atmosphere of Venus. The isotopic compositions of Kr and Xe in different volatile element reservoirs of the Solar system are plotted in Fig. \ref{fig:KryptonXenonIsotopes} and briefly discussed here to illustrate the high potential of Kr and Xe isotopic measurements to shed light on volatile origins and loss processes on Venus. 

The isotopic composition of terrestrial atmospheric krypton \textcolor{red}{is close to the Chondritic} end-member. Note that the latter two components are not simply related to one another by mass-dependent fractionation (which would appear as a rotation of the isotopic spectrum centered on $^{84}$Kr in Fig. \ref{fig:KryptonXenonIsotopes}). On Earth, mantle Kr is showing a chondritic affinity with a deficit in neutron-rich $^{86}$Kr relative to the chondritic composition whose origin remains poorly understood \citep{peron_deep-mantle_2021}. The isotopic composition of Mars atmospheric Kr has been measured in-situ \citep{conrad_situ_2016}. Mars Kr \textcolor{red}{might present} important excesses in $^{80}$Kr and $^{82}$Kr isotopes (Fig.\,\ref{fig:KryptonXenonIsotopes}) which have been attributed to the degassing from the Martian regolith of Kr isotopes produced by neutron capture reactions on bromine \citep{rao_neutron_2002}. After correction for these excesses, Mars atmospheric Kr appears to be solar-like \citep{conrad_situ_2016}. However, Mars atmospheric Kr trapped in glass phases of Martian meteorites does not exhibit the excesses measured in situ, and this discrepancy remains unexplained \citep{conrad_situ_2016,swindle_xenon_1986,avice_noble_2018}. Note that recent measurements of krypton contained in the Chassigny meteorite reveal that chondritic Kr is present in Mars's interior \citep{peron_krypton_2022}. This suggests that, contrary to the common view, delivery of chondritic volatile elements can happen \textit{before} the accretion of solar nebular gases to the Mars atmosphere.

For xenon, the isotopic composition of the Xe in the phase Q component (Q-Xe, \citet{wieler_characterisation_1992,busemann_primordial_2000-1}), which is the main carrier of noble gases in carbonaceous chondrites, is enriched in heavy isotopes and depleted in light isotopes, corresponding to a mass-dependent isotopic fractionation of about 1 \% per atomic mass unit  (or \%$.amu^{-1}$) compared to the solar composition \citep{meshik_refined_2020}. The origin of phase Q and of the isotopic difference between Q-Xe and Solar-Xe remains debated, but common explanations invoke ionization processes, causing the mass-dependent isotopic fractionation, \citep[e.g.][]{kuga_processes_2017} accompanied by addition of Xe isotopes either derived from presolar components \citep{crowther_genesis_2013} or produced during radioactive decay of I, U and Pu \citep{marrocchi_multiple_2015}.

Xenon on Earth and Mars presents a strong positive (i.e. enrichment in heavy isotopes) isotopic fractionation (3-4 $\%.u^{-1}$) relative to primordial end-members in the Solar system. Previous studies argued that Xe had been lost from Earth and Mars during early episodes of hydrodynamic escape \citep{pepin_origin_1991,pepin_origin_2002}. However, recent studies of paleo-atmospheric xenon trapped in terrestrial rocks of Archean age revealed that the isotopic fractionation of atmospheric xenon was progressive over time \citep{pujol_chondritic-like_2011,pujol_xenon_2009,avice_origin_2017,avice_evolution_2018,bekaert_archean_2018,almayrac_possible_2021} and ceased around the time of the Great Oxidation Event \citep{avice_evolution_2018,ardoin_end_2022}. \citet{zahnle_strange_2019} proposed that xenon ions escaped from the Earth's atmosphere through Coulomb interactions in an ionized hydrogen wind, which caused a progressive depletion and isotopic fractionation of atmospheric Xe on Earth. A similar process was probably operating on Mars \citep{cassata_xenon_2022} although existing data suggest that escape and isotopic fractionation of Xe ceased at least 4.2 Ga ago \citep{cassata_meteorite_2017}.

Xenon in the Earth's atmosphere also presents interesting mass-independent features. Even after correction for the mass-dependent fractionation, Earth atmospheric Xe cannot be simply attributed to Solar Xe, Chondritic Xe or a mixture of these two end-member compositions present in the Solar system. Primordial unfractionated xenon is depleted in $^{134}$Xe and $^{136}$Xe isotopes by several percents relative to Solar or Chondritic Xe. This feature has been recognized decades ago \citep{takaoka_interpretation_1972}. \textcolor{red}{Interestingly, the severe $^{134}$Xe and $^{136}$Xe anomalies on Earth relative to other bodies of the solar system correspond to a deficit in r-process Xe isotopes several orders of magnitude higher than other nucleosynthetic deviations reported for refractory nuclides measured in meteorites \citep[\textit{e.g.,}][]{kleine_non-carbonaceouscarbonaceous_2020}.} Primordial Xe for Earth atmospheric Xe has been named "U-Xe" \citep{pepin_origin_1991} and has resisted decades of investigation \citep{pepin_hunt_1994}. Recent measurements of gases emitted by the comet 67P/Churyumov-Gerasimenko revealed that cometary xenon is also depleted in $^{134}$Xe and $^{136}$Xe relative to Solar xenon \citep{marty_xenon_2017}. This depletion is even more pronounced than for U-Xe. Mass balance calculations suggest that U-Xe could result from a mixing between 78\% of Chondritic and 22\% of Cometary xenon \citep{marty_xenon_2017}. The depletion in $^{134}$Xe and $^{136}$Xe suggests that comets could be depleted in r-process nuclides and that large-scale nucleosynthetic heterogeneities persisted across the Solar system \citep{avice_xenon_2020}.

Radiogenic $^{129}$Xe produced by the decay of the short-lived radionuclide $^{129}$I (T$_{1/2}=15.7$ Myr) sheds  light on early volatile loss from planetary reservoirs. Radiogenic $^{129}$Xe excesses compared to Solar or Chondritic compositions are only generated by I/Xe fractionation within the first $\approx$100 Myr of Solar System history. On Earth, the mantle exhibits large $^{129}$Xe/$^{130}$Xe excesses compared to chondrites, generally interpreted as evidence for significant early outgassing to generate high I/Xe ratios \citep[\textit{e.g.}][]{parai_emerging_2019}, as the solubility of iodine in silicate melts exceeds that of Xe \citep{musselwhite_early_1991}. Earth's atmosphere has a low $^{129}$Xe/$^{130}$Xe ratio relative to the mantle, but exhibits $^{129}$Xe excesses compared to a mass-fractionated non-radiogenic composition; an estimated 6.8 +/- 0.3 \% of Earth's atmospheric $^{129}$Xe budget is radiogenic \citep{porcelli_models_2002}. Part of the radiogenic $^{129}$Xe in Earth's atmosphere is derived from mantle outgassing over time: samples of Archean atmosphere show a deficit in $^{129}$Xe/$^{130}$Xe compared to the modern composition after accounting for mass-dependent fractionation, providing a constraint on the rate of mantle Xe outgassing to the atmosphere, \textcolor{red}{which, at the end of the Archean, could have been a least one order of magnitude than today} \citep{avice_origin_2017,marty_geochemical_2019}. On Mars, the atmospheric radiogenic $^{129}$Xe excess relative to a fractionated non-radiogenic composition is much greater than on Earth \citep{swindle_xenon_1986,garrison_isotopic_1998} (Fig. \ref{fig:KryptonXenonIsotopes}). However, the mantle composition determined from the martian meteorite Chassigny exhibits almost no excess in $^{129}$Xe relative to chondrites \citep{ott_noble_1988,mathew_early_2001}; outgassing of the mantle reservoir sampled by Chassigny cannot explain the large radiogenic excess in $^{129}$Xe in the martian atmosphere. A distinct reservoir that experienced very strong early degassing (to generate high I/Xe within the lifetime of $^{129}$I) must exist on Mars \citep{musselwhite_early_1991}, and outgassing of this reservoir has added radiogenic $^{129}$Xe to the martian atmosphere over time \citep{mathew_early_2001,cassata_meteorite_2017}. 

Although there is no clear determination of the abundance of Xe in the Venus atmosphere, the fact that $^{131,132}$Xe isotopes may have been detected during the Venera mission (Fig. \ref{fig:IstominDetection}, \citet{istomin_venera_1983}) but that no $^{129}$Xe peak had been identified suggests that the $^{129}$Xe/$^{132}$Xe ratio is not extremely high and is likely lower than on Mars.
Taken together, Xe isotopic signatures tell a complex story of the history of a planetary atmosphere. Bulk abundance and relative ratios of primordial isotopes can be particularly diagnostic regarding the nature of accreted volatiles and the  mechanism  and  extent  of  atmospheric  loss  to  space, such as giant impact erosion or hydrodynamic escape.  The presence and abundances of radiogenic isotopes provide further insight regarding the timing and  the  degree  of early  volatile  loss  from  mantle  reservoirs  that  have  outgassed  to  the  atmosphere. Measurements of Xe in the atmosphere of Venus would thus provide potent insights into volatile origins, major evolutionary events, and transport for both the atmosphere and interior of the planet. These all have important implications for the inventory and history of other volatiles on Venus.\newline

\subsection{Stable isotopes (C, H, O, N, S)}\label{subsec:datanstableisotopes}
\subsubsection{Elemental abundances}
Carbon, hydrogen, oxygen, nitrogen and sulfur in the Venus atmosphere are mainly stored in CO$_{2}$ (96.5$\pm$0.8\,\% by vol.) for carbon and oxygen, H$_{2}$O (30$\pm$15 ppm) for hydrogen, N$_{2}$ (3.5$\pm$0.8\,\%) for nitrogen and SO$_{2}$ (150$\pm$30 ppm) for sulfur \citep[][and refs. therein]{fegley_venus_2014}.

Estimates for the elemental abundances of carbon and nitrogen in the Venus atmosphere have been reported and discussed by \citet{halliday_origins_2013}. Earth and Venus have similar abundances of carbon and nitrogen and the C/N ratio is close to chondritic proportions. Comparatively, C and N are depleted in abundances on Mars relative to chondrites, Earth and Venus, but the C/N ratio is close to the chondritic value in all three terrestrial planets. However, a near-chondritic C/N ratio should be interpreted with caution considering that strong isotopic fractionation of nitrogen isotopes in the Mars atmosphere demonstrates that Mars suffered from atmospheric escape of nitrogen \citep[][and refs. therein]{wong_isotopes_2013}. The low abundance of water in the Venus atmosphere probably results from intense episodes of hydrogen escape (see section \ref{sec:isotope_ratios} and \citet{salvador_magma_2022}).

\begin{figure*}
  \includegraphics[width=1\textwidth]{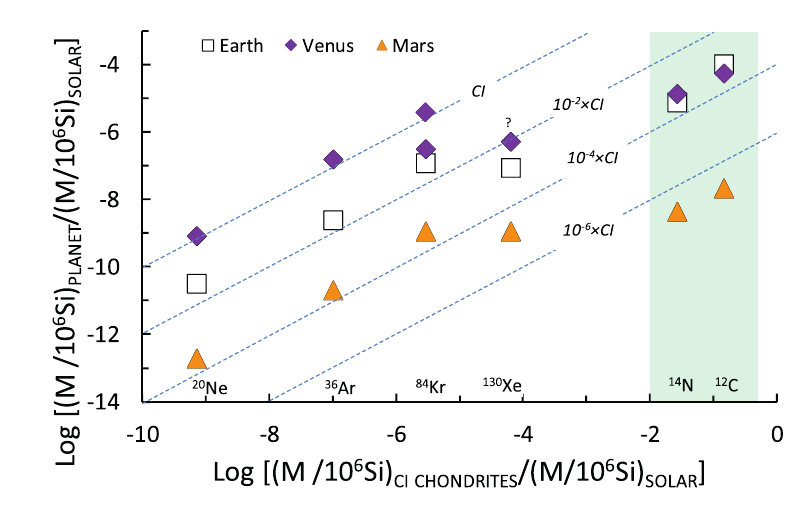}
\caption{\textcolor{red}{FIGURE MODIFIED DURING THE REVISION.} Elemental abundances of noble gases, nitrogen and carbon on bulk Earth, Venus and Mars compared to abundances in CI chondrites. Elemental abundances are normalized to the abundance of silicon and to the Solar composition. Dashed lines depict different levels of depletion compared to volatile elements in CI chondrites. The two very different estimates for the abundance of krypton are also represented (see text). See \citet{halliday_origins_2013} for details. Figure modified after \citet{halliday_origins_2013}.}
\label{fig:CN}       
\end{figure*}

\subsubsection{Isotope ratios}
\label{sec:isotope_ratios}
\begin{figure*}
\centering
  \includegraphics[width=1\textwidth]{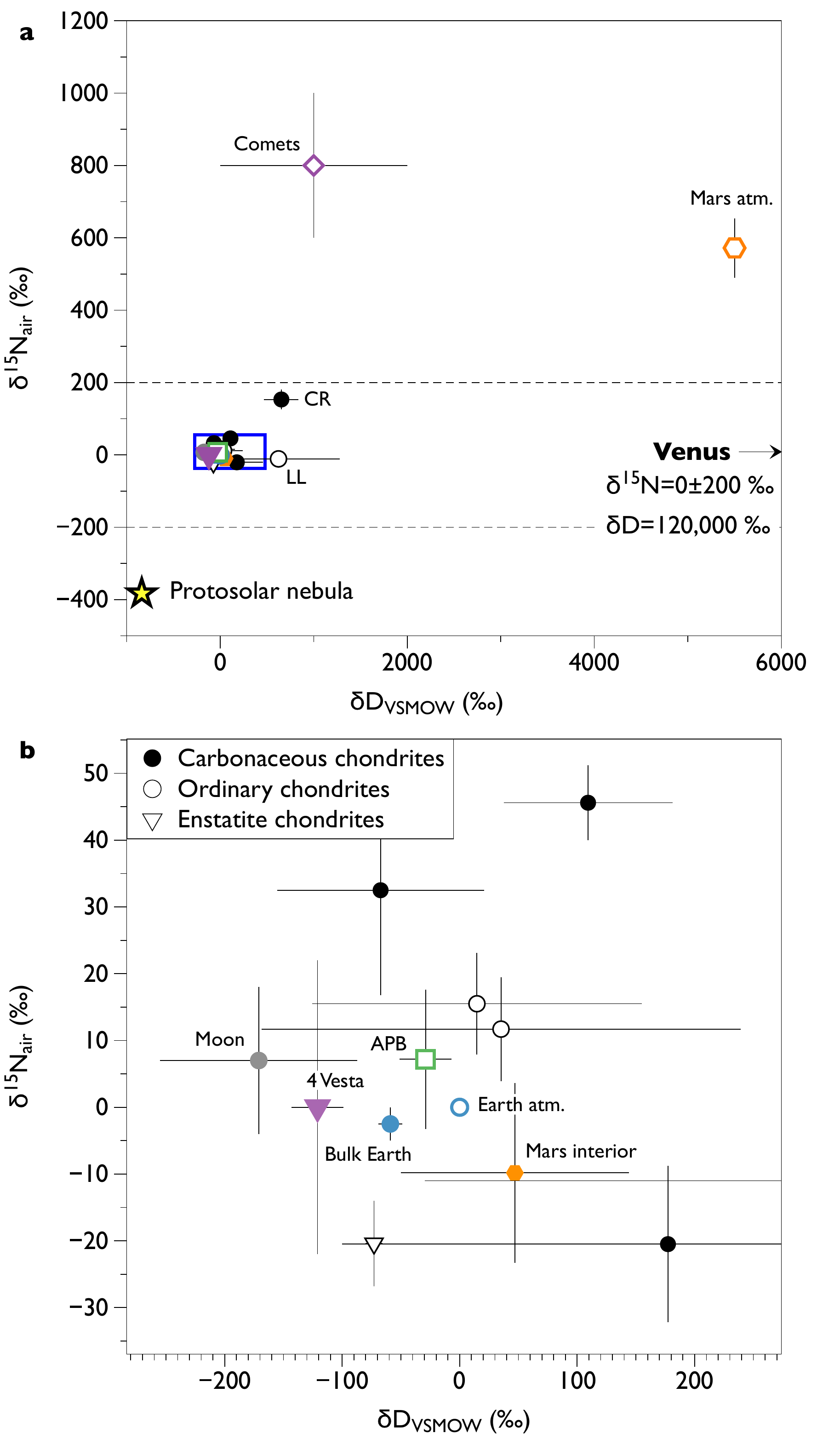}
\caption{\textcolor{red}{FIGURE MODIFIED DURING THE REVISION.} Isotopic composition of hydrogen and nitrogen in reservoirs of the solar system. (a) Venus shows an extreme $\delta$D value compared to all reservoirs of hydrogen in the solar system. The two dashed lines show the upper and lower bound for the $\delta^{15}$N value. The blue rectangle represents the zoom region for sub-panel (b). Isotope ratios are expressed with the delta notation, which corresponds to the deviation in permil relative to a standard isotope composition (mean Earth's ocean water for hydrogen, and Earth's air for nitrogen). Figure adapted from \citet{marty_origins_2012,furi_nitrogen_2015,marty_origins_2016}. Data are from \citet{mccubbin_origin_2019,marty_origins_2016} and refs. therein.}
\label{fig:DeltaD15N}       
\end{figure*}

The isotopic composition of hydrogen and nitrogen for Venus are compared to other Solar System reservoirs in Fig. \ref{fig:DeltaD15N}.
One of the most important feature of the Venus atmosphere is that the D/H ratio (where D$=^{2}$H) of hydrogen is extremely high compared to all other reservoirs in the Solar System. The $\delta$D$_{VSMOW}$ value reaches 120,000\textperthousand. This very high ratio has been interpreted as  evidence that Venus suffered from hydrogen escape episodes during its geological history \citep{donahue_venus_1982}. When hydrogen escapes from an atmosphere, deuterium is more likely to be left behind, as it is twice as heavy \citep{hunten_mass_1987}. The fraction of hydrogen remaining in the atmosphere becomes enriched in deuterium (D) relative to hydrogen ($^{1}$H). Atmospheric escape was thus an important process in the history of hydrogen on Venus. It must be noted that, since the starting D/H ratio of Venus remains unknown, an elevated $\delta$D$_{SMOW}$ also leaves room for some cometary contribution to the delivery mix of volatile elements to Venus.
For nitrogen, the $\delta^{15}$N value has been estimated at 0±200\textperthousand. This values is far too imprecise compared to typical values measured for nitrogen in the Solar system (Fig. \ref{fig:DeltaD15N}, \citet{furi_nitrogen_2015}) to draw any conclusion on the origin of Venus nitrogen and to evaluate whether the original ratio has been increased by atmospheric escape processes, as on Mars \citep{mcelroy_isotopic_1976}.

The $^{12}C/^{13}C$ ratio of carbon in $CO_{2}$ has been measured at $89.3\pm1.6$. This value is identical , within errors, to the isotopic composition of carbon on Earth \citep{istomin_mass_1980}. For oxygen, the $^{17}O/^{18}O$ ratio remains unknown and the $^{18}O/^{16}O$ value is $0.0020\pm0.0001$ \citep{hoffman_composition_1980}. This estimate is far too imprecise to evaluate if the isotopic composition of oxygen on Venus differs from other oxygen reservoirs in the Solar system \citep{greenwood_what_2020}. Finally, there is no isotopic data for sulfur, an important constituent of the atmosphere of Venus.

\section{\bf Unsolved and new questions}\label{sec:questions}
There have been no new data \textcolor{red}{acquired in-situ} on the elemental abundances and isotopic compositions of noble gases and stable isotopes in the atmosphere of Venus for the last 40 years. However, recent discoveries regarding the origin and evolution of volatile elements Earth and Mars and in other Solar System reservoirs have raised new and critical scientific questions pertaining to Venus, such that previous efforts to define the future of scientific investigations on the atmosphere of Venus \citep[e.g.][]{chassefiere_evolution_2012} must be updated. This section presents the most pressing outstanding scientific questions on the origin and early evolution of the atmosphere of Venus.

\subsection{Origin of Venus and of its atmosphere}

Noble gases and stable isotopes are often used to estimate the contributions of volatile elements from different accreted components to planetary atmospheres. In the case of Venus, current estimates of the elemental abundances and isotope ratios of those elements remain too imprecise to draw any firm conclusions, but the existing dataset allows one to pose scientific questions and frame testable hypotheses:
\begin{itemize}
    \item \textbf{To what extent did solar-derived gas contribute to the Venus atmosphere?} Although the elemental Ne/Ar ratio of the Venus atmosphere is close to the chondritic value, the $^{20}$Ne/$^{22}$Ne ratio \textcolor{red}{seems} higher than the average ratio measured in carbonaceous chondrites and closer to values measured in material irradiated by the Solar wind \citep[e.g.][for a recent review on SW noble gases]{peron_origin_2018} or to the ratio measured for solar Ne \citep{heber_noble_2009}. One possible explanation would be that Venus Ne is of solar origin and that some of this Ne escaped from the Venus atmosphere. This would lead to a increase in the Ar/Ne ratio together with a decrease of the $^{20}$Ne/$^{22}$Ne ratio. However, the existing dataset remains imprecise and a mixture between Solar Ne and Meteoritic Ne could also be envisaged (Fig. \ref{fig:NeonIsotopes}). Better measurements of the elemental abundances and the isotope ratios of Ne and Ar could shed new light on this problem. Precise measurements of the isotopic composition of Kr and Xe, \textcolor{red}{down to the percent level}, could also provide answers since the different volatile reservoirs of the Solar system (Solar gas, meteorites and even comets) have distinct Kr and Xe isotopic signatures. These end-members are not simply related by mass-dependent isotopic fractionation, which means that distinction between cosmochemical sources is possible even if the original signature has been modified by atmospheric escape leading to mass-dependent isotopic fractionation. A precise measurement of the $\delta^{15}$N value (to the 5-10\,\textperthousand\, precision level) is also desirable since the different classes of meteorites have distinct $^{15}$N/$^{14}$N ratios \citep{furi_nitrogen_2015};  
    \smallskip
    \item \textbf{Did comets contribute volatile elements to Venus?} Recent studies of cometary noble gases and comparisons with Earth's atmospheric noble gases revealed that comets may have contributed a significant portion of Earth's noble gases, while the contribution to the budget of water would have stayed relatively minor ($\sim1\%$) \citep{marty_origins_2016,marty_xenon_2017,bekaert_origin_2020}. This cometary contribution \textcolor{red}{might have left its fingerprint in the isotopic composition of Earth atmospheric xenon} \citep{avice_origin_2017,marty_xenon_2017} with a marked depletion in $^{134}$Xe and $^{136}$Xe isotopes relative to Solar or Meteoritic end-members. Contrary to the case of Earth, Mars Xe could be purely of solar origin \citep{ott_noble_1988}. Measuring precisely the isotopic composition of Venus atmospheric Xe could thus help to evaluate if comets contributed volatile elements to Venus. The Ar/Ne and $^{20}$Ne/$^{22}$Ne ratios are also of interest here. Neon condenses only at very low temperatures ($<$20K) and comets are thus probably devoid of this element \citep{bar-nun_trapping_1998}. A significant cometary contribution to the atmosphere of Venus would result in a significant shift to the right in the $^{20}$Ne/$^{22}$Ne vs $^{36}$Ar/$^{22}$Ne space (Fig. \ref{fig:NeonIsotopes}); 
    \smallskip
    \item \textbf{Was the early inner Solar system efficiently homogenized \textcolor{red}{for a major element like oxygen}?} Families of primitive meteorites have widely distinct O isotope compositions. Although the Sun represents more than 99\% of the total mass of the Solar System, meteorites and planets such as Earth and Mars are distinct from the Sun in terms of O isotope composition. Earth and Mars show enrichments in$^{17}$O up to 70\,‰ at a given $^{18}$O/$^{16}$O ratio, relative to the sun \citep{mckeegan_oxygen_2011}. This ground-breaking observation, pioneered in the 1970’s \citep{clayton_component_1973}, spurred entire fields of research. Although conflicting interpretations still exist, the current paradigm suggests the starting $\delta^{17}$O-$\delta^{18}$O composition of solids were solar, reflecting an inheritance from the average $^{16}$O-rich molecular cloud. Solids subsequently evolved towards non-solar $\delta^{17}$O-$\delta^{18}$O values via interaction with $^{17}$O-rich water: they developed positive $\Delta^{17}$O signatures relative to the Sun. Oxygen is a major mineral-forming element and makes up more than 45 wt.\% of silicate planets \citep{javoy_chemical_2010}. Scientific reasons justifying sample return from Venus for O isotope measurements were recently reviewed \citep{greenwood_what_2020}. Briefly, the later stages of terrestrial planet formation are thought to involve collisions with Moon-to-Mars sized planetesimals \citep{kaib_feeding_2015}. The final $\Delta^{17}$O signature of a planet is the result of the weighted average of the various parent bodies bodies \citep{young_oxygen_2016}. The Moon has essentially the same $\Delta^{17}$O composition as the Earth for reasons which most likely reflect its mode of formation in a giant impact event \citep{young_oxygen_2016}. In contrast, Mars and Earth have distinct oxygen isotope  $\Delta^{17}$O compositions by +0.3‰ \citep{clayton_oxygen_1983}. A variable accretionary make-up may explain differences between these two planets. Large random $\Delta^{17}$O variations may have existed among accreting bodies. Mars is only $\approx$10\% of the mass of Earth and sampled a relatively small number of accreting bodies whose weighted average $\Delta^{17}$O may have deviated from the average inner Solar System composition, whereas Earth may have inherited a composition close to the average inner Solar System, with $\Delta^{17}$O differences homogenized away by prolonged planetary growth. Alternatively, the composition of Mars may reflect an inner Solar System poorly homogenized for $\Delta^{17}$O  signatures, with a radial gradient in composition. Venus is much larger than Mars, as the second most massive terrestrial planet (after Earth) in our solar system. Whether the Earth and Venus are different in terms of $\Delta^{17}$O would be a test of the potential inhomogeneity of the inner solar system, conceivably recorded by Mars’s composition.
\end{itemize} 
    
\subsection{The problem of photochemistry}    
Photochemistry could prevent a straightforward interpretation of future O and S isotope data in terms of the origin of Venus. Processes associated with photochemistry are known to redistribute $^{17}$O between the various O-bearing molecules in the Earth's atmosphere. Ozone carries a $>$ 100 \textperthousand\, $^{17}$O anomaly in air \citep{thiemens_history_2006}. Isotope exchange between terrestrial CO$_{2}$ and ozone results in atmospheric CO2 with an anomalous $\Delta^{17}$O relative to its source \citep{thiemens_history_2006}. In fact, oxygen isotope exchange in modern air, between stratospheric ozone and any O-bearing molecule (SO$_{2}$, NO$_{x}$, CO$_2$, etc…) results in $^{17}$O signatures skewed towards anomalously positive values for all those oxygen carriers. As a result, stratospheric CO$_2$ on Earth show $^{17}$O anomalies up to ~15‰ \citep{thiemens_history_2006}, far from tracing the composition of bulk Earth. Photochemistry is known to occur in the modern atmosphere of Venus \citep[][and refs. therein]{yung_photochemistry_1982}, which is a concern for how representative Venusian CO$_2$ may be to the interior of the planet. However, the Earth mechanism described in \citet{thiemens_history_2006} is unlikely to be translated to Venus. First, a glaring difference is that there is no ozone in the Venusian atmosphere. Second, the Venusian atmospheric composition is crushingly dominated by CO$_2$, in stark contrast with Earth. For instance, the second most abundant O-bearing molecule on Venus is SO$_2$ (150 ppm), followed by water vapor (20 ppm) and carbon monoxide (17 ppm). We suggest that the mass balance is favorable for CO$_2$ to remain mostly unaffected by exotic unidentified chemistry. In other words, it is unclear how any unidentified chemistry involving third-party molecules could alter the $^{17}$O of Venus’s CO$_2$ atmosphere, as occurs on Earth.
    
Sulfur may help to understand the role of photochemistry. It occurs in the nebular gas as H$_{2}$S \citep{lauretta_experimental_1997,lodders_solar_2003}, which absorbs light and photo-dissociates when irradiated by the deep UV at wavelengths between 160 and 60 nm \citep{okabe_photochemistry_1978}. In theory, photodissociation of H$_{2}$S produces S$_{0}$ with $^{33}$S and $^{36}$S isotopic anomalies \citep{chakraborty_sulfur_2013}, like what is seen for $^{17}$O. However, differentiated meteorites show almost no $^{33}$S and $^{36}$S variations \citep{antonelli_early_2014,dottin_iii_sulfur_2015}. This is a major difference with oxygen: only extremely small $^{33}$S and $^{36}$S variations are anticipated for bulk planets. The similarity for $^{33}$S and $^{36}$S of Earth and Mars support this suggestion \citep{franz_isotopic_2014,labidi_non-chondritic_2013}. However, sulfur-bearing molecules such as SO$_{2}$ are readily photolyzed by the modern Sun’s light, in planetary atmospheres. Volcanic SO$_{2}$ on both Earth and Mars is known to develop $^{33}$S and $^{36}$S anomalies, when sent flying to optically thin regions of planetary atmospheres \citep{baroni_mass-independent_2007,gautier_2600-years_2019}. $^{33}$S and $^{36}$S measurements of Venusian SO$_{2}$ would provide first order constraints on whether any photochemistry is able to modify the composition of Venusian gases, like it does on the surface of Mars \citep{dottin_evidence_2018,franz_isotopic_2014}. These measurements will help discussing whether $^{17}$O in Venusian CO$_{2}$ is a pristine measurement of the bulk planet composition, or if atmospheric samples are inevitably skewed by the occurrence of photochemistry.

\subsection{Early evolution of the planet}

\subsubsection{Atmospheric escape and xenon}

The high D/H ratio of the Venus atmosphere has often been interpreted as evidence that Venus lost its original water \citep{donahue_venus_1982}. Dissociation of water molecules in the upper layers of the Venus atmosphere followed by escape of hydrogen to space, both powered by the strong irradiation from the early Sun, would have resulted in a depletion of water and an increase in the D/H ratio. However, both the initial water content and starting D/H ratio of water on Venus remain unknown. It remains possible that small amounts of water with a high D/H ratio \citep{altwegg_67p/churyumov-gerasimenko_2015} have been delivered to Venus by cometary bodies. Interestingly, a recent study \textcolor{red}{combined outputs from several modeling approaches (thermochemical model of convection on Venus, atmospheric escape model and N-body simulations of the formation of the solar system)} to point out that atmospheric escape on Venus was not able to remove large quantities of water over Venus history \citep{gillmann_dry_2020}. \textcolor{red}{This could imply that Venus was never affected by late accretion of water-rich material. Note however that existing models proposing scenarios for the evolution of water on Venus rely on rather poorly constrained parameters such as the initial water content or the EUV flux from the young Sun. Also, most of the isotopic fractionation of hydrogen, leading to an increase of the D/H ratio, would probably have happened only during the late stages of atmospheric escape of hydrogen. Indeed, when large amounts of hydrogen are lost during intense episodes of escape, deuterium also leaves the atmosphere efficiently and the D/H ratio stays relatively constant \citep{zahnle_mass_1990}. In other words, when the planet suffered from different atmospheric escape regimes in its history, the D/H ratio the present atmosphere brings information on only a modest part of the history of atmospheric escape.}
 
Noble gases hold clues on the extent of mass-fractionating escape processes suffered by Venus. Several models attempted to reproduce the elemental and isotopic composition of neon and argon measured in the atmosphere of Venus \citep[e.g.][]{pepin_origin_1991,gillmann_consistent_2009,lammer_constraining_2020,lammer_formation_2021}. Most models involving atmospheric escape manage to reproduce the data in a consistent way but new measurements are required to really draw firm conclusions on the history of atmospheric escape on Venus. For the time being, the existing dataset only allows to conclude that the early Sun powering atmospheric escape was either a moderate or a slow rotator \citep{lammer_constraining_2020}.

\begin{figure*}
  \includegraphics[width=1\textwidth]{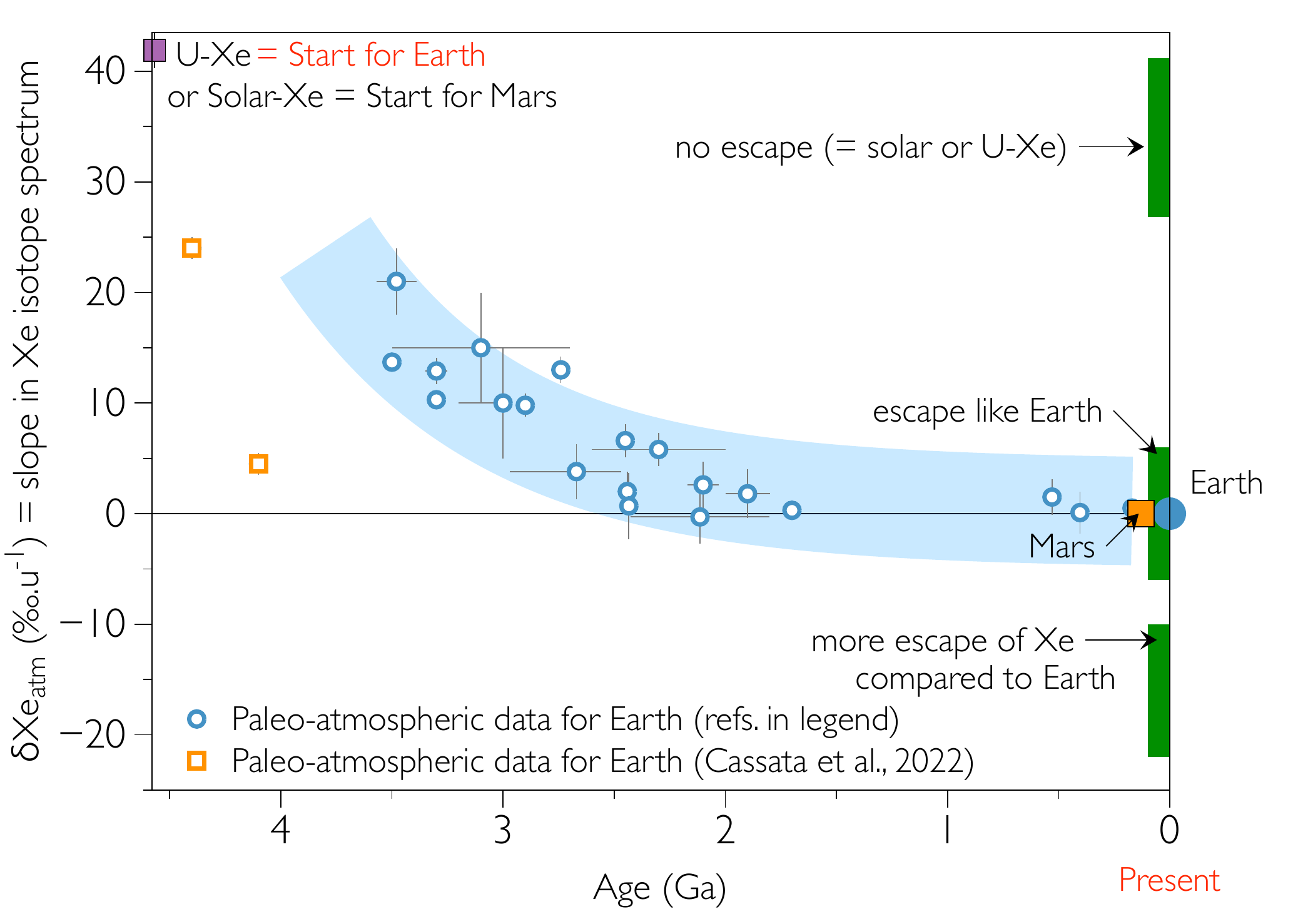}
\caption{\textcolor{red}{FIGURE MODIFIED DURING THE REVISION.} Evolution of the isotopic composition of atmospheric xenon on Earth and Mars. Isotopic fractionation is expressed relative the isotopic composition of atmospheric xenon of modern Earth and Mars. Three theoretical scenarios are displayed for Venus Xe: No escape of xenon and an isotopic composition similar to primordial xenon from Solar, Chondritic or Cometary sources; Same fractionation as for Earth and Mars; More escape (and isotopic fractionation) of xenon. Paleo-atmospheric data for Earth are from \citet{avice_perspectives_2020} and refs. therein and from \citet{ardoin_end_2022} and \citet{broadley_high_2022}. Paleo-atmospheric data for Mars are taken from \citet{cassata_xenon_2022}. Error bars are at 1$\sigma$.}
\label{fig:EvolXe}       
\end{figure*}

Recent studies of ancient terrestrial samples containing paleo-atmospheric gases revealed that the isotopic composition of terrestrial atmospheric Xe evolved during at least 2 Ga after the Earth formed \citep{pujol_chondritic-like_2011,pujol_xenon_2009,avice_origin_2017,avice_evolution_2018,bekaert_archean_2018,ardoin_end_2022}(Fig. \ref{fig:EvolXe}). This evolution corresponds to a progressive mass-dependent fractionation of Xe isotopes probably starting from a composition corresponding to a mixture between cometary and meteoritic Xe \citep{marty_xenon_2017} and reaching the modern-like isotopic composition around the time of the Great Oxidation Event ca. 2.3 Ga ago \citep{avice_evolution_2018,ardoin_end_2022}. The current favored explanation for explaining this protracted evolution is an escape of xenon ions together with hydrogen ions from the Archean atmosphere to outer space \citep{zahnle_strange_2019,catling_archean_2020}. The isotopic composition of atmospheric Xe could thus be a tracer of hydrogen escape from planetary atmospheres. Compared to Earth, xenon in the atmosphere of Mars is likely derived from a different source, \textit{i.e.} solar gas \citep{garrison_isotopic_1998,ott_noble_1988} but also presents a mass-dependent fractionation of 3-4 \%$.u^{-1}$ relative to the starting isotopic composition \citep{swindle_xenon_1986}. The magnitude of fractionation is similar to the case of Earth atmospheric xenon. Assuming that atmospheric escape of Xe is responsible for both the depletion and the isotopic fractionation of the remaining fraction, this similarity in the extent of isotopic fractionation is intriguing given that the the escape process will be governed by parameters (abundance of total H$_{2}$, irradiation from the Sun, magnetic field etc.) which are likely to have been very distinct for early Earth and Mars. Note that, although present-day isotopic fractionation of atmospheric xenon is similar for Earth and Mars, the fractionation (and maybe escape) of xenon on Mars probably ceased much earlier than for Earth, around 4.2 Ga ago \citep{cassata_meteorite_2017,cassata_xenon_2022} (Fig. \ref{fig:EvolXe}). Collecting data on the abundance and isotopic composition of Xe in the Venus atmosphere is a high priority to determine to what extent "missing and isotopically fractionated xenon" is a common feature of planetary atmospheres of the inner Solar system and to further elucidate if atmospheric escape played a role in shaping the elemental and isotopic composition of atmospheric Xe. \textcolor{red}{Note that potential fissiogenic additions to the budget of atmospheric $^{131-136}$Xe isotopes pose a challenge when one wants to determine the fractionation of atmospheric xenon \citep[\textit{e.g.},][]{avice_origin_2017}. A precise determination of the abundances of less abundant light isotopes such as $^{124,126,128,130}$Xe plays then an important role to estimate the degree of fractionation of atmospheric xenon.}

Further, the isotopic composition of nitrogen may also shed light on atmospheric loss in Venus’ history.  It is a reasonable hypothesis that the terrestrial planets formed with a similar nitrogen source, and the $^{15}$N/$^{14}$N ratio places an important constraint on if and when Venus could have lost its putative primordial ocean and on the timing of the runaway greenhouse onset. Nitrogen could have been preferentially lost from Venus’ atmosphere largely during its early geologic history when the atmosphere was presumably relatively thin \citep[][and refs. therein]{baines_atmospheres_2013}. Once the atmosphere became massive and CO$_2$-rich, further fractionation of isotopes in bulk nitrogen may be less evident due to dilution of N$_2$ by CO$_2$ at the exobase. Therefore, if the $^{15}$N/$^{14}$N ratio is significantly greater than the terrestrial value (thus closer to Mars), this would suggest that intense hydrodynamic escape must have occurred prior to the rise of CO$_2$.

\subsubsection{Contributions from mantle outgassing and characterization of mantle reservoirs}
It is of vital importance to understand the role of mantle outgassing in shaping surface conditions on Venus \citep{gillmann_long-term_2022}. Identification of radiogenic $^{129}$Xe excesses in the atmosphere of Venus would suggest that at least some portion of the interior experienced early degassing to fractionate I/Xe during the lifetime of $^{129}$I, and has subsequently outgassed to affect the Venus atmospheric composition. The magnitude of the radiogenic excess, if any, would provide a lower limit constraint on early outgassing of a Venus mantle reservoir for comparison with terrestrial and martian mantle reservoirs, yielding insights into comparative geodynamics during accretion, and could allow comparisons of the degree of outgassing to the atmosphere. If no radiogenic $^{129}$Xe excess is observed in the atmosphere, this would suggest either that the Venus mantle did not experience early outgassing (similar to the Chassigny mantle source on Mars), or that outgassing of the mantle has been very limited over the past 4.45 Ga, pushing any catastrophic early outgassing to explain low atmospheric $^{40}$Ar/$^{36}$Ar to the earliest stages of Venus history. Xe isotopic measurements of the Venus atmosphere are needed to understand the timing and extent of mantle outgassing, and the relationship between early outgassing and atmospheric loss to space \citep{cassata_meteorite_2017,cassata_xenon_2022}. \textcolor{red}{Note that atmospheric erosion by impacts could also have altered the budget of radiogenic xenon, depending on the timings of outgassing relative to the impacts.}. Combined, abundances of the radiogenic noble gases $^{129}$Xe, $^{40}$Ar, and $^{4}$He will contribute to the understanding of the relative rates and significance of early, long-term, and recent outgassing, respectively.

\section{\bf Recommendations for future investigations}
The existing dataset on the abundance and isotopic composition of noble gases and stable isotopes in the atmosphere of Venus is partial and imprecise and new investigations are urgent.
A list of key measurements of noble gases and of their associated maximal uncertainties required to answer the scientific questions described in Section \ref{sec:questions} is summarized in Table 1.

Two broad types of science investigation could be envisaged for gathering data on the elemental and isotopic compositions of noble gases and stable isotopes (H, C, N, O, S) in the atmosphere of Venus. One type would be an in-situ mission carrying a scientific payload able to measure the abundances and isotope ratios of the chemical elements of interest. Another one would be a sample return mission during which a portion of the Venus atmosphere would be sampled. The collected sample(s) would then be returned to Earth for characterization with state-of-the-art technologies available in international laboratories. The pros and cons regarding measurements of noble gases and stable isotopes are presented and briefly discussed in the next sections. See also Chapter 4 of this issue for a detailed discussion of future missions as well as new concepts for exploring Venus.

\subsection{In situ measurements}
\begin{figure*}
  \includegraphics[width=0.75\textwidth]{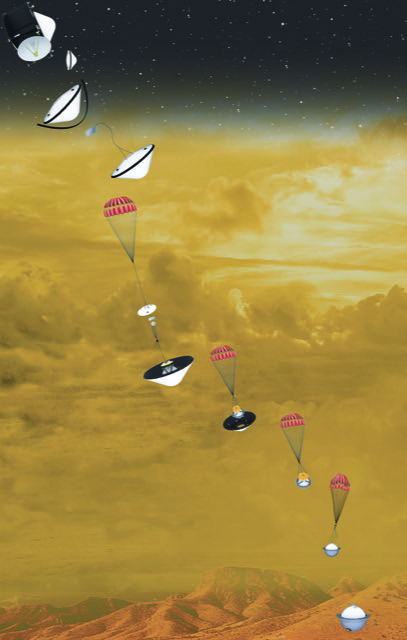}
\caption{The NASA DAVINCI mission \citep{garvin_revealing_2022} will conduct in situ sampling of the Venus atmosphere during its $\approx$hour-long descent to the surface, providing critical measurements of noble gases and stable isotopes.}
\label{fig:DAVINCI}       
\end{figure*}

To date, almost all data on the abundance and isotopic composition of volatile elements in the atmosphere of Venus \citep{johnson_venus_2019} have been collected during in situ investigations by probes plunging through the atmosphere of Venus and carrying mass spectrometers  \citep[\textit{e.g.}][]{ehrenfreund_mass_1999}. This is the approach that will be used by the recently-selected DAVINCI mission from NASA \citep[][Fig. \ref{fig:DAVINCI}]{garvin_revealing_2022}. 

One major advantage of in situ investigations is that the sample of interest is taken directly from the well-mixed atmosphere around the probe.  In  situ  investigations  at  low  velocity  and  well below  the  homopause  ensure  that  the  measured  gas  is  representative  of the  entire  atmosphere and that detected abundances and ratios are unambiguous \citep{tian_atmospheric_2015}. Deep atmosphere sampling also avoids any uncertainties introduced by temporal or spatial variability of the homopause location by species or other upper atmosphere stratification \citep{von_zahn_upper_1980,mahieux_densities_2012,gruchola_upper_2019-1,peplowski_chemically_2020}.

Further, in situ sampling greatly minimizes complications of terrestrial contamination, as instrumentation can be cleaned and sealed off prior to launch, particularly for the volatiles. The captured in situ sample will thus greatly exceed any residual terrestrial species and dominate the measured signal. Finally, measurement of the atmosphere during a descent allows for sampling at multiple altitudes for species where there is the potential for vertical distribution that would otherwise complicate interpretation. Variable measurements of the D/H ratio at Venus suggest there may be a gradient \citep{bertaux_warm_2007}, potentially driven by photolysis-induced isotopic fractionation \citep{liang_modeling_2009}, preferential escape, or selective condensation, a process that has been found to be important for fractionating D and H on Mars and Earth \citep{bertaux_isotopic_2001}.

 In  situ  investigations do present important challenges to be mitigated by each mission.  Mass  spectrometers  on  board  atmospheric  probes  are inherently  limited  compared  to  ground-based  analytical  abilities, due to necessary constraints in terms of power consumption, size, mass, measurement duration, and robustness against spaceflight environments \citep{arevalo_mass_2020}. Three key considerations for mass spectrometric measurements of noble gases in particular are sensitivity, selectivity, and dynamic range. Sensitivity and resolution (as a proxy for selectivity) are two commonly limiting factors for flight mass spectrometers and become extremely important when targeted species are rare isotopes (D, $^{3}$He, $^{78,80}$Kr, $^{124-128}$Xe, $^{33}$S etc.). The instrument must be sufficiently sensitive to detect a signal corresponding to these rare species. The dynamic range must also be sufficiently large to accurately measure ratios of species whose abundances span orders of magnitude (i.e., Xe vs. Ar). Another complication  stems  from  the  fact  that  isobaric  interferences  are  always  present, either in the residual background of the instrument itself or in the sample, even after purification steps  \citep{wieler_noble_2014}. For example, for some ionization energies and instruments, doubly ionized CO$_{2}^{++}$ ions are detected at mass 21.9949 while $^{22}$Ne$^{+}$ ions are detected at mass 21.9914. The mass resolving power of the instrument required to simultaneously measure pure signals of each species if both are present in the instrument is higher than 6,000. Similarly, H$^{35}$Cl and H$^{37}$Cl signals are isobaric interferences of $^{36}$Ar and $^{38}$Ar, respectively (mass resolving power of 3,000). \textcolor{red}{In the absence of high resolution for in situ instruments, the ionization energy can be adjusted to minimize the prevalence of doubly-charged ions. For other interferences, additional preprocessing and gas-cleaning techniques are required.} 
 
 One way these three challenges have been addressed in previous in situ investigations has been to include an enrichment system to process samples prior to introduction into the instrument \citep{niemann_galileo_1992,mahaffy_sample_2012}. Isobaric interferences can be minimized by significantly increasing the  noble  gas  to  interference  ratio  by  working  in  ultra-clean  conditions and by employing efficient purification methods to remove the bulk atmospheric components. Gas enrichment units have been employed  (\textit{e.g.} the SAM experiment onboard the Curiosity rover, \citet{franz_initial_2017}) to artificially increase the partial pressure of noble gases, successfully enabling the measurement of Xe isotopes during a descent at Jupiter \citep{mahaffy_noble_2000} as well as recently in situ Ar, N$_2$, Kr, and Xe isotopic ratios from the surface of Mars \citep{atreya_primordial_2013,wong_isotopes_2013}. \textcolor{red}{Note that for Mars data, the N$_{2}$/Ar ratio was subsequently corrected after taking into account results obtained with calibration cells \citep{franz_initial_2017}. The corrected N$_{2}$/Ar ratio is now in agreement with results obtained by previous space missions \citep{owen_composition_1977} and by analyses of SNC meteorites \citep{avice_noble_2018}.}
 Performing enrichments in a step-wise manner can also address isobaric interferences across targeted species as well expand the effective dynamic range. The improved capabilities are necessarily traded against increases in complexity,  size,  power  consumption, and measurement duration of the experiments.

 Measurements of stable isotopes are also challenged by the presence of isobaric  interferences within a mass spectrometer (\textit{e.g.} $^{12}$C$^{16}$O signal interfering with $^{28}$N$_{2}$ signal). Additional steps of chromatography can partially solve this problem since it allows to introduce the chemical species at distinct times into the mass spectrometer. Alternatively, unlike noble gases, many of the volatile species that carry isotopes of interest have strong spectral features and thus can be measured with high sensitivity and specificity using Tunable Laser Spectroscopy (TLS) \citep{tarsitano_multilaser_2007}. Isotopic ratios of carbon, oxygen, and hydrogen have been measured using TLS on Mars \citep{franz_indigenous_2020}, and similar technique could be adapted to in situ measurements at Venus even on a descent probe.
 
 Finally, given the harsh conditions encountered by a probe plunging in the Venus atmosphere, all the scientific tasks must be achieved very rapidly, on the order of the hour, and the scientific data must be transferred either to an orbiting spacecraft or directly to Earth as soon as possible. Long-lived surface platforms, or aerial platforms at more temperate altitudes, could enable repeatable measurements with longer processing and integration times. However, these are less technically mature, more costly and the discussion of their inherent challenges is beyond the scope of this contribution.

\subsection{Sample return}
There are many scientific motivations for mission concepts proposing to return a sample from the Venus atmosphere and several mission scenarios have already been envisaged \citep[\textit{e.g.}][]{rodgers_venus_2000,sweetser_venus_2003,greenwood_oxygen_2018}. In most cases, the sequence of atmospheric sampling is part of a larger complex mission involving a descent stage, ground operations by a lander etc. One technical challenge is then to find a method to launch the atmospheric samples from Venus ground to orbit \citep[\textit{e.g.}][]{rodgers_venus_2000}.
One alternative scenario for a sample return mission scenario would be a probe launched from Earth, plunging at high velocity through the Venus atmosphere, sampling the atmosphere  \citep{sotin_cupids_2018} and returning the sample back to Earth. Such a mission can be achieved with a free-return ballistic trajectory \citep{sweetser_venus_2003}. One important advantage of this type of scenario is that the scientific payload for this type of mission could be very simple with only few instruments to characterize the sampled gas such as pressure gauges. Most of the payload would consist in the sampling system (pipes, valves and sampling cylinders). This simple approach would certainly save mission costs and reduce the risk of this type of mission. Importantly, there would be no stringent constraint on the delay between sampling and transfer in new containers followed by scientific analyses if the sampling cylinders present very low leak rates and very low degassing rates of unwanted background gases (\textit{e.g.} H$_{2}$) that could compromise the original chemical composition of the sample. Finally, and this is probably the most relevant argument here, sampled gas could be measured on Earth with state-of-the-art mass spectrometry \& spectroscopy techniques, some of them being non-destructive such as Tunable Laser Spectroscopy (TLS) \citep[\textit{e.g.}][]{crosson_stable_2002,tarsitano_multilaser_2007}. This non-destructive technique would allow to make a thoughtful characterization of the sample with, for example, measurements of the isotope ratio of H, O, C. High precision noble gas mass spectrometry could be done on very small amounts of the collected gas \citep{wieler_noble_2014}, especially if high-sensitive methods are used such as resonance ionization mass spectrometry \citep{gilmour_relax:_1994}. 
Although many of the scientific questions raised in Section \ref{sec:questions} could be, at least partially, answered with an \textit{in-situ} sampling and characterization mission, returning an atmospheric sample from Venus would bring the scientific output of any mission to Venus to another level. For example, determination of the $\Delta^{15}N^{15}N$ value of Venus atmospheric nitrogen, \textcolor{red}{which quantifies the excess in the $^{15}$N$^{15}$N molecule relative to a random distribution of N atoms in N$_{2}$,} would allow to investigate the photochemistry of atmospheric nitrogen on Venus and to evaluate the amount of exchanges between Venus interior and its atmosphere \citep{yeung_extreme_2017,labidi_hydrothermal_2020,gillmann_long-term_2022}. Similarly, applying very sensitive isotope determination techniques to a sample from the Venus atmosphere would allow detection and precise measurement of the abundances of minor isotopes of noble gases ($^{3}$He but also $^{78}$Kr and $^{124,126}$Xe) in order to better evaluate the delivered mix of volatile elements to the Venus atmosphere but also to put constraints on atmospheric escape processes or on interior-surface interactions.

However, a sample return mission faces important challenges. In the case of a free-return ballistic trajectory, the sample will be collected at high velocity ($>10\,km.s^{-1}$) and the gas in the front of the probe will be shocked, brought to high temperatures, and eventually turned into a plasma \citep{sweetser_venus_2003}. Gas will then be fed into the sampling system via an inlet port and will have to travel through pipes and valves until a final expansion step into the sampling cylinder. All the steps described above could alter the chemical and isotopic integrity of the sample. The presence of a plasma implies that molecules will be dissociated. Cooling of the plasma will eventually lead to a recombination of the atoms into new molecules not originally present in the atmosphere of Venus.
Noble gas atoms will also be ionized in the plasma and will be implanted on the surfaces of the probe. Numerous studies pointed out that the isotopic composition of noble gas trapped in the upper layers of solid exposed to ion implantation is mass-dependently fractionated relative to the starting composition \citep{bernatowicz_isotopic_1987,kuga_processes_2017}. Transfer of the gas from the atmosphere to the cylinder could also induce chemical and/or isotopic fractionation of gaseous species depending on the parameters of the gas flow. The effects of high velocity sampling on the chemical and isotopic composition of the sampled gas are currently studied via numerical simulations using a Direct Simulation Monte Carlo (DSMC) approach \citep{rabinovitch_hypervelocity_2019} and analog experiments for high velocity sampling are currently under development.

Returning an atmospheric sample from Venus requires important technical developments for the sampling system and the sample containers, and a full curation procedure for the sample(s) must be carefully prepared. Technologies for storing gas samples have already been employed in previous space missions \citep{allen_curating_2011,moeller_sampling_2021}. A sample container must fit important criteria including: (i) a proper seal with a helium leak-rate on the order of 10$^{-10}$ scc/s or below \citep{moeller_sampling_2021}; (ii) minimal interactions between sampled gas and container material; (iii) a sufficiently robust mechanical structure to resist Earth return methods (capture on orbit or direct re-entry). Sample curation is an extremely complex and costly task \citep[\textit{e.g.}][]{mccubbin_advanced_2019}. Furthermore, curation of an atmospheric sample from another planet would be a novelty and would rest on very different principles compared to most existing setups for curating extraterrestrial rocks. For example, part of the modern sample curation process for samples recovered from asteroid Itokawa by the japanese \textit{Hayabusa} mission (JAXA) involves high-vacuum chambers and the use of pure gases in order to preserve the sample from any terrestrial contamination \citep{yada_hayabusa-returned_2014}. Preserving and handling a gas sample from the Venus atmosphere would, of course, rely on different principles (no direct pumping nor gas injection), which remain to be defined by the international community.

Interestingly, a mission to Venus returning atmospheric samples to Earth would likely fit in the "Unrestricted Category V" of the classification scheme established by the Committee on Space Research \citep{cospar_cospar_2020}. If the classification is not revisited in the future, it would mean that such a sample would be subject to less stringent restrictions regarding its potential bearing of traces of life compared to samples from Mars, Europa or Enceladus. This would certainly relax the constraints for sample handling, preliminary characterization and preparation for long-term curation. However, some studies pointed out that Venus may have presented habitable surface conditions until as little as 700 Ma ago \citep{way_was_2016} and that life forms could be hosted in the lower cloud layer on present-day Venus \citep{limaye_venus_2018}. See also Westall et al. (this issue). 
It is plausible that, in the near future, Venus will join Mars, Europa and Enceladus in the list of objects for which a special care must be taken if a sample return is planned.

\section{\bf Conclusions}
Knowing the elemental abundances and isotopic compositions of noble gases and H, C, N, O, S in the Venus atmosphere enables setting important constraints on the origin and early evolution of Earth's sister planet. Existing data are incomplete, but important differences compared to Earth and Mars have already been pointed out by previous studies. Venus is enriched in light noble gases compared to Earth and Mars. The Ar/Ne ratio is close to chondritic values whereas the isotopic composition of neon may be solar-like. There is little radiogenic $^{40}$Ar in the atmosphere of Venus compared to Earth, suggesting that Venus is less degassed compared to Earth. Data on the abundance and isotope composition of krypton and xenon are cruelly lacking; these data are needed to put new constraints on the delivery mix of volatile elements to the entire planet and on the age and history of the atmosphere. For stable isotopes, the very high D/H ratio suggests that Venus suffered from hydrogen escape although the starting abundance and isotopic composition of Venus water remains unknown.
A precise determination of the isotopic composition of nitrogen could help constrain the extent of elemental and isotopic fractionation of hydrogen and nitrogen by thermal and non-thermal atmospheric escape processes. Measurements of the isotopic composition of xenon will determine if, like Earth and Mars, Venus suffered from coupled hydrogen-xenon escape processes.
About future investigations, the scientific community and institutional decision makers can opt for two types of investigations offering interesting complimentary perspectives: in-situ measurements or sample return. There is a strong and useful heritage of space technologies designed for in-situ measurements with an impressive list of outstanding scientific results obtained on diverse objects in the Solar System. Base on this approach, the recently selected DAVINCI (NASA) mission will provide a wealth of new data on the elemental and isotopic composition of noble gases and stable isotopes contained in the atmosphere of Venus.
Sample return concepts present important but surmountable technical, scientific and organisational challenges. It may be within reach in the coming decades. Returning a sample from the atmosphere of Venus would provide, with a careful curation of the returned sample, enough analysable material to achieve high-precision measurements and would certainly lead to scientific breakthroughs.

%
%
\label{tab:1}       

\begin{acknowledgements}
G.A. acknowledges the french Centre National d'Etudes Spatiales (CNES) for its funding support for Venus related studies. 
We warmly thank Tilman Spohn and members of the team at ISSI (International Space Science Institute) in Bern for their hospitality. We also thank the two anonymous reviewers for their comments and the editors for their careful handling of the manuscript.
\end{acknowledgements}

%
 \section*{Conflict of interest}
 The authors declare that they have no conflict of interest.

\bibliographystyle{spbasic}      
\bibliography{BIB.bib}   


%

\end{document}